\documentclass{aa}

\usepackage{graphicx}
\usepackage{txfonts}
\usepackage{lscape}
\usepackage{hyperref}
\usepackage{xcolor}
\usepackage{lscape}

\newcommand\Halpha{\mbox{${\mathrm H} \alpha$}}%
\newcommand\Hbeta{\mbox{${\mathrm H} \beta$}}%
\newcommand\NII{[\ion{N}{II}]}%
\newcommand{\ngc}[1]{NGC~{#1}}%
\newcommand{\acs}[1]{ACS~ID~{#1}}%
\newcommand\CVSGGC{CVSGGC}

\newcommand\nfh{\mbox{$^{\mathrm h}$}}%
\newcommand\nfm{\mbox{$^{\mathrm m}$}}%

\newcommand{\changed}[1]{\textcolor{black}{#1}}%

\newcommand\nxray{39}%
\newcommand\nemstars{156}%

\begin{document}

   \title{A stellar census in globular clusters with MUSE: A spectral catalogue of emission-line sources}
   \titlerunning{A spectral catalogue of emission-line sources in Galactic globular clusters}

   \author{Fabian G\"ottgens \inst{1}
          \and
          Tim-Oliver Husser \inst{1}
          \and
          Sebastian Kamann \inst{2}
          \and
          Stefan Dreizler \inst{1}
          \and 
          Benjamin Giesers \inst{1}
          \and
          Wolfram Kollatschny \inst{1}
          \and 
          Peter M. Weilbacher \inst{3}
          \and
          Martin M. Roth \inst{3}
          \and
          Martin Wendt \inst{4}
          }

   \institute{Institut für Astrophysik, Georg-August-Universit\"at G\"ottingen, Friedrich-Hund-Platz 1, 37077 G\"ottingen, Germany\\
              \email{fabian.goettgens@uni-goettingen.de}
         \and
             Astrophysics Research Institute, Liverpool John Moores University, 146 Brownlow Hill, Liverpool L3 5RF, United Kingdom 
         \and 
	     Leibniz-Institut für Astrophysik Potsdam (AIP), An der Sternwarte 16, 14482 Potsdam, Germany 
	 \and 
	     Institut für Physik und Astronomie, Universität Potsdam, Karl-Liebknecht-Str. 24/25, 14476 Golm, Germany 
	 }
   \date{Received XXX; accepted YYY}

  \abstract
   {}
   {Globular clusters produce many exotic stars due to a much higher frequency of dynamical interactions in their dense stellar environments.
   Some of these objects were observed together with several hundred thousands other stars in our MUSE survey of 26 Galactic globular clusters. 
   Assuming that at least a few exotic stars have exotic spectra, that means spectra that contain emission lines, we can use this large spectroscopic data set of over a million stellar spectra 
   as a blind survey to detect stellar exotica in globular clusters.}
   {To detect emission lines in each spectrum, we model the expected shape of an emission line as a Gaussian curve.
   This template is used for matched filtering on the differences between each observed 1D spectrum and its fitted spectral model.
   The spectra with the most significant detections of \Halpha{} emission are checked visually and cross-matched with published catalogues.}
   {We find \nemstars{} stars with \Halpha{} emission, including several known cataclysmic variables (CV) and two new CVs, 
   pulsating variable stars, eclipsing binary stars, the optical counterpart of a known black hole, several probable sub-subgiants and red stragglers, and 21 background emission-line galaxies. 
   We find possible optical counterparts to {\nxray} X-ray sources, as we detect {\Halpha} emission in several spectra of stars that are close to known positions of Chandra X-ray sources. 
   This spectral catalogue can be used to supplement existing or future X-ray or radio observations with spectra of potential optical counterparts to classify the sources.} 
   {}


   \maketitle
%
\section{Introduction}

In the dense stellar environments of globular clusters (GCs), the frequent interactions between stars produce a wealth of stellar exotica.
This includes interacting binary systems and end states of stellar evolution, such as cataclysmic variables \citep[CVs, ][]{ivanova_formation_2006}, pulsars \citep{ransom_pulsars_2007}, 
and planetary nebulae (PNe) \citep{jacoby_masses_2017}.
Emission lines are expected to appear in the optical spectra in at least some of these stellar types; those stars are then classified as emission-line stars.
Because of the old age of globular clusters and their stars, some types of emission-line stars still present in the Milky Way disc do not exist (anymore) in globular clusters, for example Wolf-Rayet stars or Be stars.

In recent years, several stellar-mass BH candidates have been found in binary systems in globular clusters \citep{strader_two_2012,giesers_detached_2018}. 
Photometric observations suggest that some of these systems could be {\Halpha} emitters. 
While stellar-mass BHs were long thought to be ejected from GCs during cluster evolution, 
the discoveries of stellar-mass BH candidates in multiple clusters indicate a large population of these black holes inside evolved GCs \citep{strader_two_2012,askar_mocca-survey_2018,kremer_how_2018}.

CVs are binary systems consisting of a hot, compact white dwarf and a dwarf star in a close orbit. 
The white dwarf accretes material from its companion star that accumulates in an accretion disc.
In the dense stellar environments of globular clusters, CVs and progenitor systems are influenced by dynamical interactions, 
with up to 50\,\% forming via a binary encounter \citep[][but also see \citealt{belloni_mocca-survey_2019}]{ivanova_formation_2006}.
The number of predicted CVs per cluster is on the order of 200, but the number of observed CV candidates or confirmed CVs in the literature is much lower \citep{knigge_cataclysmic_2012}.
CV candidates can be found with photometric observations, for example looking for dwarf nova outbursts, for stars with UV excess \citep[e.g.][]{sandoval_new_2018}, 
for outliers in the colour-magnitude diagram of a GC \citep[e.g. ][]{campos_outliers:_2018},
or using {\Halpha} surveys \citep{knigge_cataclysmic_2012}.
Alternatives for detecting CVs are far-ultraviolet spectroscopy which has also been useful for detecting CVs in globular clusters \citep{knigge_far-ultraviolet_2003}, and X-ray observations.
Follow-up optical spectroscopy to confirm CVs in GCs is difficult because of the crowded fields and the intrinsically low brightness of CVs.

When a nova occurs in a CV, it can leave behind a visible emission nebula as a remnant such as the one in \ngc{6656} \citep{gottgens_discovery_2019}. 
Nova remnants are not the only type of nebula in GCs, another type are planetary nebulae of which four are known in the Galactic GC system.
Even this low number of PNe in GCs is too high because the low masses of AGB stars should prohibit the formation of PNe \citep{jacoby_planetary_1997}.
This lead to the prediction that PNe in GCs are formed by a different mechanism, possibly by binary interaction \citep{jacoby_planetary_1997,jacoby_masses_2017}.

Many stellar exotica in GCs have been found using X-ray observations. However, it is less clear which optical counterpart belongs to an X-ray source when only broad-band photometry is available.
In this case, a counterpart is identified if it is an outlier in the optical CMD with respect to all other cluster stars, 
that is if its separation from the main sequence or the red-giant branch is too large, or if its color is too blue \citep[e. g. ][]{bassa_x-ray_2004,webb_x-ray_2004}.
Similarly, the an optical emission lines in a spectrum of a star close to an X-ray source could also indicate it is a counterpart.

Previous optical surveys for typical classes of emission-line objects used photometric observations and the on/off-band technique, variability, or anomalous colours to detect candidate objects:
\citet{jacoby_planetary_1997} conducted the most successful PNe survey for GCs, \citet{knigge_cataclysmic_2012} lists several CV surveys.
Spectroscopic follow-up observations are then used to confirm the classification and to derive more properties of the source.

The data used in this paper were obtained with MUSE \citep{bacon_muse_2010}, a panoramic integral-field spectrograph at the Very Large Telescope, as part of a survey of Galactic globular clusters.
With the MUSE data already obtained, emission-line objects can be found without the need of additional observations because both spatial and spectral information is present. 
While \citet{roth_muse_2018} demonstrated the efficiency of MUSE at detecting emission-line objects including Wolf-Rayet stars, 
supernova remnants, \ion{H}{ii} regions, and PNe in the galaxy \ngc{300},
we can for the first time conduct a blind survey for emission-line stars in Galactic globular clusters.

\section{Data}
\subsection{Observations and reduction}
\label{sct:data_obs}

\begingroup
\renewcommand{\arraystretch}{1.2} 
\begin{table}
\caption{Overview of globular cluster data used in this paper. This includes all observations made between September 2014 and \changed{March 2019.}}
\label{table:overview_obs}
\centering
\tiny
\begin{tabular}{rcrrrrr}
\hline
  NGC & Name & $N_{\rm pointings}$ & $N_{\rm epochs}$ & ToT [h] & $N_{\rm spectra}$ & $N_{\rm stars}$ \\
 (1) & (2) & (3) & (4) & (5) & (6) & (7) \\ \hline
 104 & 47\,Tuc & 10 &   13 & 12.2 & 309911 & 32055 \\
 362 &  & 6 &    2 &  1.3 & 24049 & 9363 \\
 1851 &  & 4 &  6.5 &  4.7 & 67267 & 11614 \\
 1904 & M\,79 & 4 &  5.5 &  2.5 & 32597 & 5669 \\
 2808 &  & 4 &    2 &  1.2 & 20230 & 8040 \\
 3201 &  & 5 &   12 & 11.6 & 61855 & 4503 \\
 5139 & $\omega$\,Cen & 10 & 10.5 & 12.2 & 335614 & 45616 \\
 5286 &  & 1 &    4 &  1.0 & 17954 & 8282 \\
 5904 & M\,5 & 6 &    2 &  2.3 & 51450 & 18203 \\
 6093 & M\,80 & 4 &    2 &  1.6 & 21051 & 9153 \\
 6121 & M\,4 & 2 &    1 &  0.1 & 1251 & 1067 \\
 6218 &  & 4 &    3 &  3.0 & 22989 & 6616 \\
 6254 & M\,10 & 8 &  1.5 &  3.3 & 29633 & 14296 \\
 6266 & M\,62 & 4 &    3 &  2.0 & 39190 & 15900 \\
 6293 &  & 1 &    2 &  0.1 & 2154 & 1326 \\
 6388 &  & 4 &    4 &  2.2 & 46600 & 14484 \\
 6441 &  & 4 &    4 &  2.8 & 43473 & 13247 \\
 6522 &  & 1 &    3 &  0.2 & 7564 & 3567 \\
 6541 &  & 5 &    2 &  1.9 & 35352 & 12003 \\
 6624 &  & 1 &    2 &  0.5 & 8300 & 4556 \\
 6656 & M\,22 & 4 &  2.5 &  2.2 & 36609 & 13204 \\
 6681 & M\,70 & 1 &    2 &  0.8 & 8283 & 4773 \\
 6752 &  & 8 &    2 &  3.0 & 31070 & 14086 \\
 7078 & M\,15 & 4 &    3 &  1.6 & 40606 & 18015 \\
 7089 & M\,2 & 4 &    4 &  2.4 & 47764 & 15309 \\
 7099 & M\,30 & 4 &  3.5 &  2.6 & 34176 & 9111 \\
  &  &  &  &  &  \\
 {\bf total} &  & 114 & 103.0 &  80.8 & 1379362 & 316428 \\
\hline\end{tabular}
\medskip
\tablefoot{(1) NGC number. (2) Alternative identifier (if any). (3) Number of pointings. This number roughly corresponds to the covered field of view in arcminutes. 
(4) Average number of epochs available for each pointing. (5) Total integration time in hours. (6) Number of extracted spectra.  (7) Number of stars with at least one extracted spectrum.}

\end{table}
\endgroup

This work makes use of all data taken with MUSE for our survey of 26 Galactic globular clusters between September 2014 and March 2019 (PI: S.~Dreizler, S.~Kamann).%
\footnote{ESO Program IDs: 094.D-0142, 095.D-0629, 096.D-0175, 097.D-0295, 098.D-0148, 099.D-0019, 0100.D-0161, 0101.D-0268, and 0102.D-0270.}
MUSE has a large field of view ($1\arcmin \times 1\arcmin$) combined with a spatial sampling of {0.2\arcsec} and an intermediate resolution $R$ between 1800 and 3500 in the spectral range from 4750 to 9350~\AA.
The observations and the analysis steps have been described in detail by \citet{kamann_stellar_2018} and are summarised here.
In contrast to \citet{kamann_stellar_2018}, this work also includes data from observations made after October 2016.
Table~\ref{table:overview_obs} gives an overview of the observation statistics for each cluster.

Each observation was reduced with the standard MUSE pipeline \citep{weilbacher_design_2012,weilbacher_muse_2014} which calibrates the images from the 24 MUSE spectrographs, 
including cosmic ray rejection, and transforms them into a datacube.
In the next step, single stellar spectra are extracted with a point-spread-function (PSF) from this datacube.
The extractions use the PSF-fitting developed in \citet{kamann_resolving_2013} to measure the PSF parameters and determine stellar positions in the datacube as a function of wavelength.
We mostly use stellar positions from the ACS survey of Galactic globular clusters \citep[][hereafter ACS catalogue]{sarajedini_acs_2007,anderson_acs_2008} as an input for the extraction, see Table~2 in \citet{kamann_stellar_2018} for details.
The extracted spectra are then analysed with respect to the Göttingen spectral library \citep{husser_new_2013}, a grid of synthetic spectral models suitable for most stars in globular clusters.
A chi-square fit on the full spectrum minimises the difference between the observed spectrum and a model spectrum 
by interpolating between grid spectra
to determine the stellar parameters effective temperature, metallicity, and the radial velocity \citep{husser_muse_2016}.

In addition to the spectra obtained from single observations, we also use these to create a high-signal-to-noise spectrum for each star.
We shift all spectra of each star to the Sun's restframe and add the flux weighted by the signal-to-noise ratio of the spectrum.
These combined spectra are then analysed in a similar way to the one described above. 

\subsection{Residuals from spectral fitting}
\label{sct:residuals}
The residuals from the spectral fitting are defined as the difference between model and observation. 
We can use these residuals to detect emission-line stars because the spectral library does not contain spectra with emission lines, in other words
if an emission line is present in the observed spectrum, it will also be visible in the residuals.

The residuals can contain random noise, additional absorption from the interstellar medium \citep{wendt_mapping_2017}, 
systematic errors of the model spectra (e.g.\ absorption lines that only exist in the models, see Fig.~\ref{fig:matched_explanation}, or vice versa), instrumental systematics, and true emission lines.
If the fit did not find the global minimum of the chi-square space, the residuals will contain a large amount of stellar light.
In this case, the parameters determined by the fit do not necessarily describe the star, and the residuals can cause false positive detections of emission lines. 

It is also possible that the spectral model grid does not contain a suitable model for the observed spectrum.
This occurs for horizontal branch stars and some M stars. Spectra of M stars contain strong molecular bands which have a great influence on the overall spectral appearance.
A slight mismatch in the fit of an M star spectrum has a large impact on the residuals.
If a spectrum of an M star contains emission lines, the effect of the emission lines on the residuals could be smaller than the effects of spectrum mismatch.
In these cases, the method based on matched filtering to detect emission lines described in Sect.~\ref{sct:matched_filter} is not reliable and the detection fails.
However, the method based on the residuals without convolution (Sect.~\ref{sct:plain_residuals}) still works in these cases.

\section{Emission line detection}

As shown in Table~\ref{table:overview_obs}, we extract millions of stellar spectra from our observations. 
The large number of observed spectra makes it impossible to visually check each of them. 

We use two approaches to detect emission lines in the residuals from the spectral fit. 
The first approach based on matched filtering is widely applied to similar problems, 
for example in gravitational-wave detection \citep{abbott_observation_2016} and to detect emission-line galaxies in MUSE datacubes \citep{herenz_lsdcat:_2017}.
The second method uses only the residuals and a running estimate of the noise. 
This method is used as a backup whenever matched filtering fails to detect a signal, because it is much simpler and more robust but also produces more false detections.
Both approaches assign a significance to each detection which is then used to select the most promising candidates for visual inspection.

We stress that we do not use existing catalogues of emission-line stars as a prior to find those in our data.
Since the aim of this work is to find new and unexpected sources, we use external catalogues only after our methods identify a possible spectrum with emission lines.

\subsection{Matched filtering with mean absolute deviation}
\label{sct:matched_filter}

   \begin{figure}
   \centering
      \includegraphics[width=0.5\textwidth]{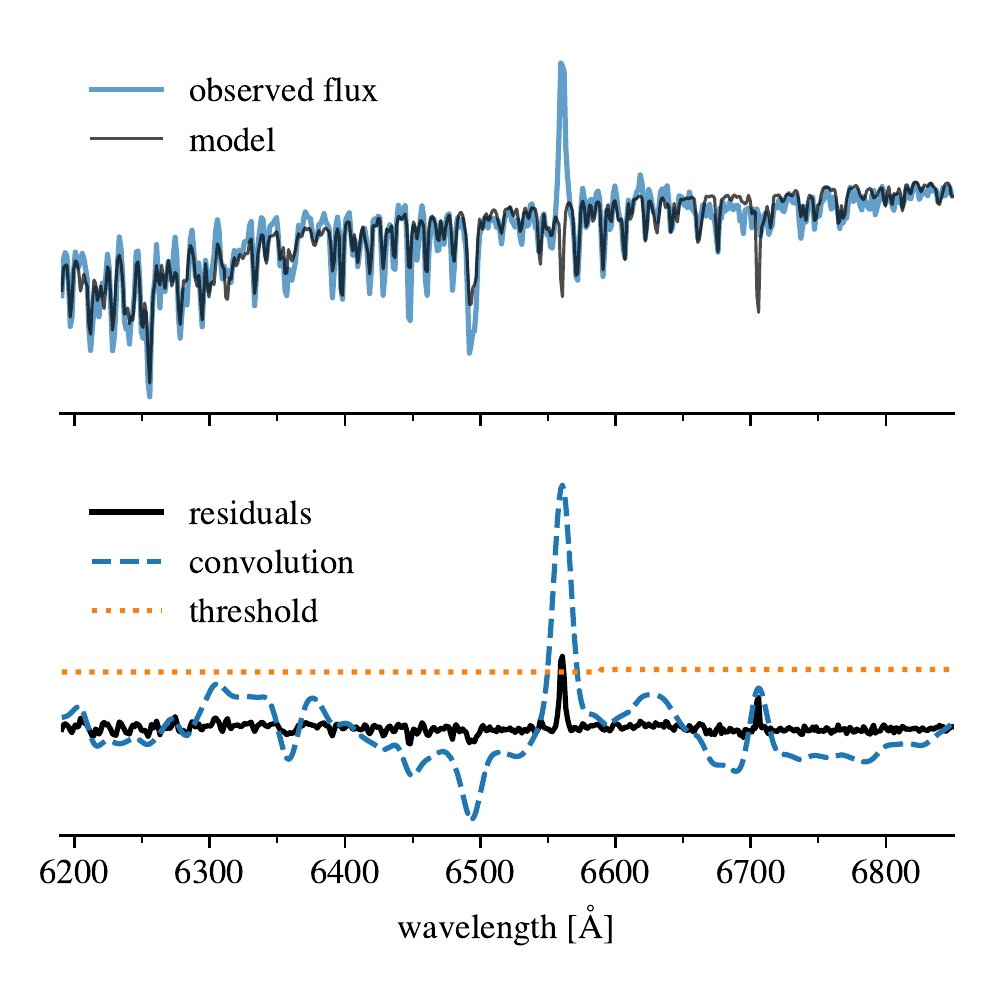}
      \caption{Illustration of emission-line detection with matched filtering. Top: observed flux of V190 in \ngc{6266} with {\Halpha} emission and the fitted spectral model. 
      Bottom: The {\Halpha} emission line is still present in the residuals, as well as an absorption line at 6700~{\AA} which is only found in the models. 
      The convolution with a Gaussian increases the emission line above the threshold calculated from the mean absolute deviation of the residuals.
      }
         \label{fig:matched_explanation}
   \end{figure}
   
Because of the large dataset, we need to chose an approach that is fast and can extract potentially weak signals.
One algorithm with these properties is called \textit{matched filtering} \citep[see][and references therein]{vio_correct_2016} that requires prior knowledge about the expected signal.
We assume that each emission line can be described by a Gaussian curve with a standard deviation (width) of 5~{\AA}. This width is determined from simulations in Sect.~\ref{sct:efficiency}.
Instead of applying matched filtering directly on the observed spectral flux (top panel in Fig.~\ref{fig:matched_explanation}), 
we use it on the residuals that result from spectral fitting (see bottom panel of Fig.~\ref{fig:matched_explanation}).
Mathematically, matched filtering computes the convolution $C(\lambda)$ of the filter (the expected line profile) and the residual flux.
As shown in the bottom panel in Fig.~\ref{fig:matched_explanation} (dashed line), the convolution is high at a certain wavelength when the expected flux shape matches the measured flux.
On the other hand, noise -- which is typically only a few pixels wide -- is smoothed out, that means the convolution gives a much lower value as it would for real emission.
Compared to the convolution of noise or continuum flux with a Gaussian, emission flux appears in the convolution as a peak centred at the emission line.
We detect an emission signal at wavelength $\lambda_e$ if the convolution at that point is larger than some threshold function $t(\lambda)$ at the same point.
The threshold function is constructed as the median absolute deviation (MAD) calculated separately for wavelength bins of the residual flux (dotted line in the bottom panel of Fig.~\ref{fig:matched_explanation}).
By construction, the ratio $D_s = C(\lambda_e)/t(\lambda_e)$ is higher for more prominent emission signals.
We call this ratio detection significance and use it to select promising candidates for visual inspection by requiring that a detection lies above a minimum value of $D_s$.
The detection efficiency depends on this choice and it is analysed with simulated emission lines in Sect.~\ref{sct:efficiency}.

\subsection{Plain residuals and running noise estimate}
\label{sct:plain_residuals}
This section presents a more robust method of detecting emission lines that relies on fewer assumptions. 
Similar to the method based on matched filtering, it relies on the residuals from the spectral fit.
The residual flux $r_i = r(\lambda_i)$ at each wavelength point $\lambda_i$ is compared to the residual noise $s_i$ at the adjacent wavelengths.
We locally estimate the noise from the difference of the ninetieth and tenth percentile of the residual flux in a 100~{\AA} window centred at $\lambda_i$.
Since the spectral model does not describe the observed flux perfectly, the residuals contain noise and systematic effects (see Sect. \ref{sct:residuals}).
We account for these outliers in the residual flux by using percentiles instead of extrema or measures that are sensitive to outliers.
At each wavelength, the ratio of residual flux to the noise estimate $D_s = r_i/s_i$ represents the significance of an emission-line detection.
For comparison, if the noise was normally distributed with a variance $\sigma^2$, a ratio of $D_s = 1$ corresponds to an observation with a significance of $\approx 1.3 \sigma.$

Simulations show that this method works well for narrow emission lines but not for broad ones. 
This is because a broad emission line increases the residuals over a broader spectral range, and thus the percentiles used to estimate the noise increase as well.
Since this increases $s_i$ but not the amplitude $r_i$, the detection significance $D_s$ decreases accordingly.

\subsection{Detection efficiency}
\label{sct:efficiency}

   \begin{figure}
   \centering
      \includegraphics[width=0.5 \textwidth]{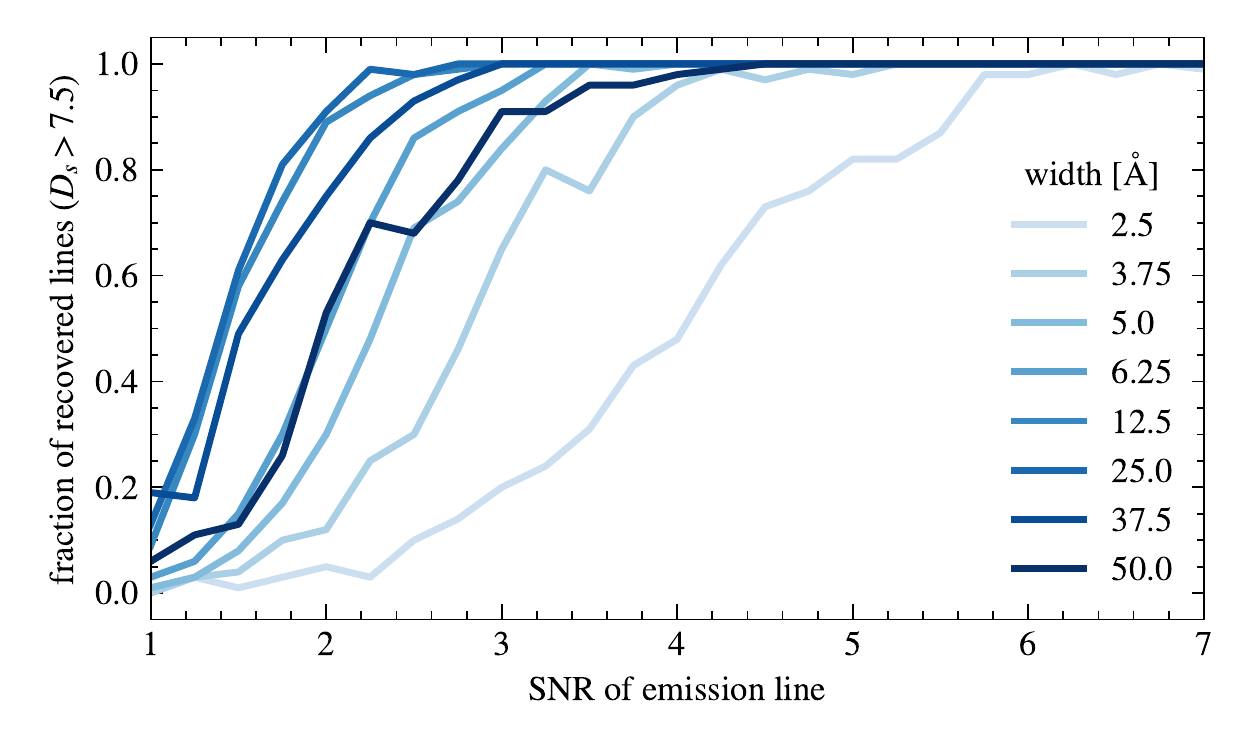}
      \caption{Fraction of simulated emission lines that are recovered with a filter width of $6~\AA$ depending on the simulated line width. 
               Only detections above a detection significance $D_s > 7.5$ are taken into account.
              }
         \label{fig:detected_fraction}
   \end{figure}

We estimate the fraction of emission signals recovered with the method based on matched filtering described in Sect.~\ref{sct:matched_filter} with simulated emission lines.
The detection significance of an emission-line candidate depends on the amplitude and width of the emission peak in the residual flux,
and on the noise of the spectrum and the width of the Gaussian filter. 
We construct emission lines by sampling a Gaussian curve with a standard deviation $\sigma$ that we vary between 3 and $60\,\AA$, its amplitude is set to one.
We draw noise from a normal distribution with a width of $(SNR)^{-1}$ and add it to the signal.
For each simulated emission line, we apply the detection method (Sect.~\ref{sct:matched_filter}) and calculate the detection significance for filter widths of 6, 25, 60, and 120\,\AA.
We note that this leads to an estimate of how effective the detection methods is with respect to the amplitude of an emission line and not the total line flux.

The main results of this analysis are: 
We can find broad synthetic emission lines of several tens of \AA ngstr\"om even with a narrow filter width of $6\,\AA$ while the reverse is not true. 
As Fig.~\ref{fig:detected_fraction} shows, we recover about 50\,\% of broad emission lines ($\approx 40\,\AA$) with a filter width of 6\,\AA for a SNR of 2
if we only take detections with $D_s > 7.5$ into account. 
This fraction increases with increasing SNR and with decreasing emission-line width (except for very narrow widths $< 5\,\AA$ below the filter width).
We conclude from these simulated emission lines that for our data the choice of a 6~{\AA} filter is reasonable, and $D_s > 7.5$ is a good lower limit for a detection to be inspected further. 
The filter width is four to five times the FWHM of the line-spread-function (LSF) of MUSE which varies between 2.5 and 3~{\AA} depending on wavelength \citep{bacon_muse_2017}. 
In principle, one could choose the threshold $D_s$ much lower than this for the price of many more detections to inspect, which will contain a much higher frequency of false positives.
The choice of $D_s > 7.5$ is also justified by the low empirical true positive rate of $\lesssim5\%$ below this limit (see Sect.~\ref{sct:results} and Fig.~\ref{fig:true_positive_rate}).

\subsection{Limiting flux}

\begingroup
\renewcommand{\arraystretch}{1.5} 
\begin{table}
\caption{Limiting fluxes for all clusters, calculated using a broad emission line (40~\AA) and also valid for narrow lines (see text).
Uncertainties give the central 80\,\% of the distribution on the main-sequence (MS), main-sequence turn-off (TO), and red giant branch (RGB).}
\label{table:minimum_fluxes}
\centering
\begin{tabular}{lrrr}
\hline
Cluster & \multicolumn{3}{c}{$\log_{10} F_\text{min} [\text{erg/s/cm$^2$/$\AA$}]$ } \\
 & MS & TO & RGB \\
\hline
\ngc{104} &$ -16.9 \substack{ + 0.3 \\ - 0.5 } $ &$ -16.7 \substack{ + 0.3 \\ - 0.5 } $ &$ -16.3 \substack{ + 0.4 \\ - 0.2 } $ \\
\ngc{1851} &$ -17.5 \substack{ + 0.3 \\ - 0.2 } $ &$ -17.3 \substack{ + 0.3 \\ - 0.3 } $ &$ -17.0 \substack{ + 0.5 \\ - 0.3 } $ \\
\ngc{1904} &$ -17.5 \substack{ + 0.3 \\ - 0.2 } $ &$ -17.4 \substack{ + 0.3 \\ - 0.2 } $ &$ -16.9 \substack{ + 0.4 \\ - 0.2 } $ \\
\ngc{2808} &$ -17.4 \substack{ + 0.3 \\ - 0.2 } $ &$ -17.3 \substack{ + 0.3 \\ - 0.2 } $ &$ -16.8 \substack{ + 0.4 \\ - 0.2 } $ \\
\ngc{3201} &$ -17.6 \substack{ + 0.3 \\ - 0.2 } $ &$ -17.4 \substack{ + 0.3 \\ - 0.1 } $ &$ -16.6 \substack{ + 0.1 \\ - 0.1 } $ \\
\ngc{362} &$ -17.1 \substack{ + 0.3 \\ - 0.2 } $ &$ -17.1 \substack{ + 0.3 \\ - 0.2 } $ &$ -16.5 \substack{ + 0.5 \\ - 0.3 } $ \\
\ngc{5139} &$ -17.1 \substack{ + 0.3 \\ - 0.2 } $ &$ -16.9 \substack{ + 0.3 \\ - 0.2 } $ &$ -16.5 \substack{ + 0.2 \\ - 0.2 } $ \\
\ngc{5286} &$ -17.6 \substack{ + 0.3 \\ - 0.2 } $ &$ -17.5 \substack{ + 0.3 \\ - 0.2 } $ &$ -17.2 \substack{ + 0.4 \\ - 0.2 } $ \\
\ngc{5904} &$ -17.4 \substack{ + 0.3 \\ - 0.3 } $ &$ -17.2 \substack{ + 0.3 \\ - 0.3 } $ &$ -16.6 \substack{ + 0.4 \\ - 0.1 } $ \\
\ngc{6093} &$ -17.5 \substack{ + 0.2 \\ - 0.2 } $ &$ -17.5 \substack{ + 0.3 \\ - 0.2 } $ &$ -17.1 \substack{ + 0.4 \\ - 0.3 } $ \\
\ngc{6121} &$ -16.9 \substack{ + 0.3 \\ - 0.1 } $ &$ -16.9 \substack{ + 0.2 \\ - 0.1 } $ &$ -16.4 \substack{ + 0.0 \\ - 0.0 } $ \\
\ngc{6218} &$ -17.8 \substack{ + 0.3 \\ - 0.1 } $ &$ -17.4 \substack{ + 0.2 \\ - 0.1 } $ &$ -16.9 \substack{ + 0.3 \\ - 0.1 } $ \\
\ngc{6254} &$ -17.6 \substack{ + 0.3 \\ - 0.3 } $ &$ -17.4 \substack{ + 0.3 \\ - 0.2 } $ &$ -16.7 \substack{ + 0.3 \\ - 0.1 } $ \\
\ngc{6266} &$ -17.6 \substack{ + 0.3 \\ - 0.2 } $ &$ -17.4 \substack{ + 0.4 \\ - 0.2 } $ &$ -16.9 \substack{ + 0.5 \\ - 0.2 } $ \\
\ngc{6293} &$ -17.1 \substack{ + 0.2 \\ - 0.2 } $ &$ -17.2 \substack{ + 0.2 \\ - 0.2 } $ &$ -16.8 \substack{ + 0.4 \\ - 0.2 } $ \\
\ngc{6388} &$ -17.5 \substack{ + 0.2 \\ - 0.2 } $ &$ -17.4 \substack{ + 0.3 \\ - 0.2 } $ &$ -17.1 \substack{ + 0.4 \\ - 0.2 } $ \\
\ngc{6441} &$ -17.6 \substack{ + 0.2 \\ - 0.1 } $ &$ -17.5 \substack{ + 0.3 \\ - 0.2 } $ &$ -17.2 \substack{ + 0.3 \\ - 0.2 } $ \\
\ngc{6522} &$ -17.4 \substack{ + 0.3 \\ - 0.1 } $ &$ -17.4 \substack{ + 0.3 \\ - 0.1 } $ &$ -17.0 \substack{ + 0.3 \\ - 0.2 } $ \\
\ngc{6541} &$ -17.4 \substack{ + 0.3 \\ - 0.2 } $ &$ -17.3 \substack{ + 0.3 \\ - 0.2 } $ &$ -16.9 \substack{ + 0.3 \\ - 0.1 } $ \\
\ngc{6624} &$ -17.6 \substack{ + 0.3 \\ - 0.2 } $ &$ -17.6 \substack{ + 0.3 \\ - 0.2 } $ &$ -17.2 \substack{ + 0.5 \\ - 0.2 } $ \\
\ngc{6656} &$ -17.2 \substack{ + 0.3 \\ - 0.3 } $ &$ -17.1 \substack{ + 0.4 \\ - 0.3 } $ &$ -16.5 \substack{ + 0.3 \\ - 0.1 } $ \\
\ngc{6681} &$ -17.8 \substack{ + 0.3 \\ - 0.2 } $ &$ -17.7 \substack{ + 0.3 \\ - 0.2 } $ &$ -17.2 \substack{ + 0.1 \\ - 0.1 } $ \\
\ngc{6752} &$ -17.3 \substack{ + 0.3 \\ - 0.4 } $ &$ -17.0 \substack{ + 0.4 \\ - 0.3 } $ &$ -16.6 \substack{ + 0.3 \\ - 0.2 } $ \\
\ngc{7078} &$ -17.5 \substack{ + 0.4 \\ - 0.2 } $ &$ -17.4 \substack{ + 0.4 \\ - 0.3 } $ &$ -16.8 \substack{ + 0.5 \\ - 0.3 } $ \\
\ngc{7089} &$ -17.4 \substack{ + 0.3 \\ - 0.2 } $ &$ -17.3 \substack{ + 0.3 \\ - 0.2 } $ &$ -16.9 \substack{ + 0.3 \\ - 0.2 } $ \\
\ngc{7099} &$ -17.6 \substack{ + 0.3 \\ - 0.2 } $ &$ -17.4 \substack{ + 0.3 \\ - 0.2 } $ &$ -16.9 \substack{ + 0.3 \\ - 0.2 } $ \\
\hline
\end{tabular}
\end{table}
\endgroup

   \begin{figure}
   \centering
      \includegraphics[width=0.5\textwidth]{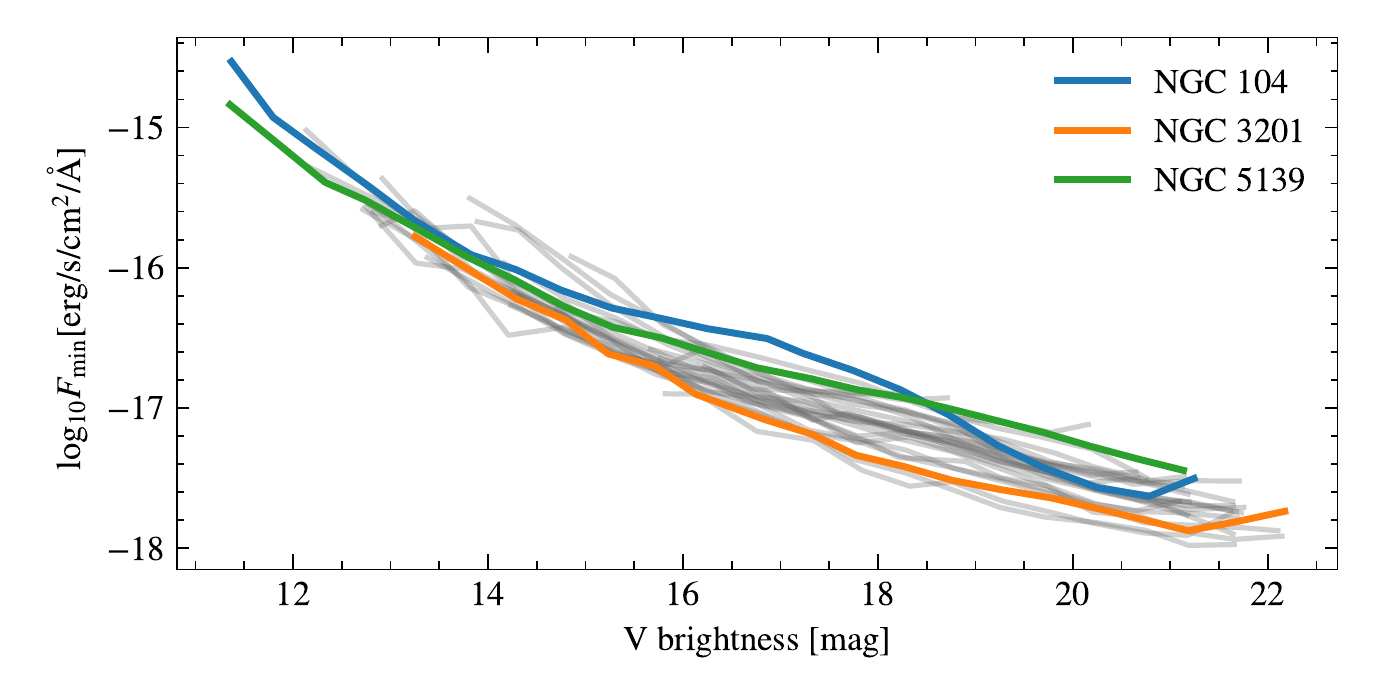}
      \caption{Limiting flux for which 80\,\% of sources will be detected as determined from simulated emission lines for different clusters as a function of brightness. 
      Each grey curve represents a GC, the clusters \ngc{104}, \ngc{3201}, and \ngc{5139} are highlighted.}
         \label{fig:limiting_flux}
   \end{figure}
   
The results of the simulations above are now used to calculate the minimum amplitude of an emission line that can be detected. 
Since we select detections with a significance above 7.5, emission lines in spectra with too high noise will not be found. 
From the simulations, we first estimate the minimum signal-to-noise ratio $\text{SNR}_\text{min}$ for that 80\,\% of simulated emission lines are found. 
This $\text{SNR}_\text{min}$ depends on the width of the simulated emission line. Here, we chose a width of 40\,{\AA}, corresponding to cataclysmic variables.
The minimum signal that we can detect is estimated by measuring the noise $\sigma$ in the residuals of all spectra from 6000~{\AA} to 7000~{\AA}.
In practice, the noise depends on the brightness of the target star, observing conditions, stellar crowding, etc. 
We measure this effective noise in the residuals obtained from the spectral fitting.
The minimum detectable signal in each spectrum is $F_\text{min} = \text{SNR}_\text{min} \cdot \sigma$. 
Table~\ref{table:minimum_fluxes} lists $F_\text{min}$ of a broad emission line for different representative points in the stellar population of each cluster we observed,
and Fig.~\ref{fig:limiting_flux} shows the limiting flux as a function of stellar brightness in three clusters.
Since we use a $\text{SNR}_\text{min}$ for which 80\,\% of simulated emission lines are found, Table~\ref{table:minimum_fluxes} gives the limiting flux for which 80\,\% of all spectra with 
an emission line are found.
Because we use a narrow filter width of 5~{\AA} to detect emission lines, the limits given for a broad emission line can be treated as a conservative estimate of the limiting flux of narrow emission lines.
The limiting fluxes for narrow emission lines generally fall inside the uncertainties given in Table~\ref{table:minimum_fluxes}, this means that this table is also valid for narrow lines.
Depending on the brightness of the target star, we find that $F_\text{min}$ is generally between $10^{-17} \ldots 10^{-16} \, \text{erg/s/cm$^2$/$\AA$}.$
To our knowledge, this is the first optical emission-line survey estimating the upper limit of fluxes for sources that remain undetected.

\section{The catalogue of emission-line sources}
\subsection{Results of visual inspection}
\label{sct:results}

   \begin{figure}
   \centering
      \includegraphics[width=0.5\textwidth]{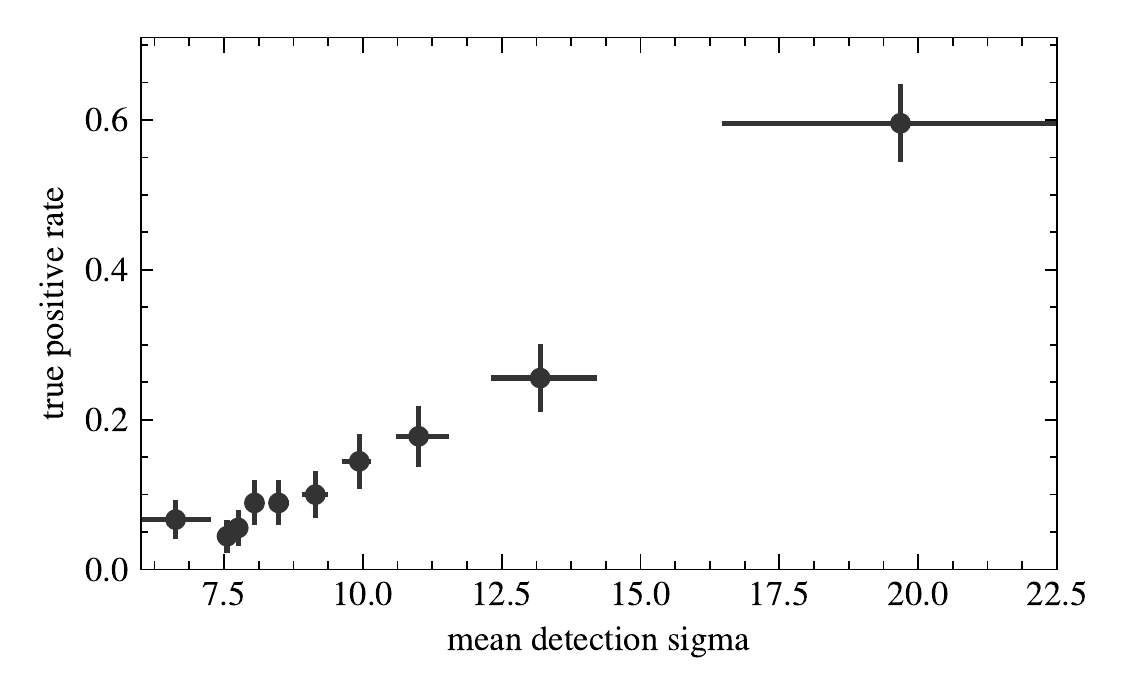}
      \caption{Empirical true positive rate of matched filtering after checking about 1200 stars with detected emission lines. 
      Each bin contains 90 stars and the errorbars in x-direction contain the central $1\sigma$ interval of the detection significance per bin.
              }
         \label{fig:true_positive_rate}
   \end{figure}
   
We applied both detection methods to spectra extracted from MUSE observations of globular clusters (see Sect.~\ref{sct:data_obs}) that have a signal-to-noise-ratio of at least five.
Setting this threshold ensures that the spectral fit gives meaningful residuals.
We want to inspect promising emission-line spectra visually. 
As this would be very time consuming with thousands of candidate spectra, 
we check candidate spectra which contain an emission line close to {\Halpha} (between 6540 and 6580~\AA) if $D_s > 7.5$.
With this set-up, we expect to find emission-line stars, typically {\Halpha} emitters, while galaxies would remain undetected.
Section~\ref{sct:galaxies} describes how we find galaxies using all detected emission lines.
In total, 1200 individual stars have at least one such spectrum, with a total of about 9000 spectra.

For each spectrum, we check if the emission line could be valid according to a set of criteria.
The potential emission line has to be at least two pixels wide (a pixel corresponds to 1.25~{\AA}) and it must fulfill at least one of the following criteria:
\begin{itemize}
\item The line candidate appears in roughly the same position with the same shape in multiple spectra of the same star, or
\item the spectrum shows emission lines in addition to {\Halpha}, or
\item the corresponding star is listed in the \textit{Catalogue of Variable Stars in Galactic Globular Clusters} \citep[CVSGGC,][]{clement_variable_2001, clement_catalogue_2017}, Simbad \citep{wenger_simbad_2000}, or in a suitable catalogue in Vizier \citep{ochsenbein_vizier_2000}, or
\item the star is close to an X-ray source as listed in the Chandra Source Catalog Release 2.0 \citep{evans_chandra_2010}.
\end{itemize}

Typically, an emission-line candidate is not valid if the spectrum seems to be contaminated by other stars or nebulae.
This occurs if a much brighter star is close ($\approx 2 \arcsec$ or less) to the target star, 
or if it is close to one of the three nebulae in our survey.

Inspection of the results show that false positives are mainly caused by noise, contamination by brighter stars, and poor fit results.
Figure~\ref{fig:true_positive_rate} shows the empirical true positive rate after a visual check of each star with $D_s > 7.5$. 
For testing purposes, we also checked emission-line candidates with a lower significance than 7.5, these stars are also included in this figure. 
As expected, the true positive rate correlates with the mean detection significance and reaches about 60\,\% for $D_s > 6.$

Table~\ref{tbl:elos} lists all stars with spectra containing valid emission lines that we found in our survey. 
This table also gives the original ID used in the ACS catalogue in column `ACS ID'.
The columns `$d_C$' lists the projected distance to the respective cluster centre.
The table also contains our estimate whether the star is a likely cluster member in column `mem.?'. In contrast to \citet{kamann_stellar_2018}, this estimate is based on radial velocities only.
Column `vrad?' contains an indicator whether the star shows variations in its radial velocity based on the method described in \citet{giesers_stellar_2019}.
We converted the probability of variability calculated in \citet{giesers_stellar_2019} in the following way: $p<0.15$: not variable, $p>0.85$: variable, $0.15<p<0.86$: unsure (?). Blank fields indicate insufficient data. We expect a false positive rate of 15\,\%.
Cross-matches with other catalogues and papers are given in column `Ident.' with the corresponding reference in column `Ref.'.
The column `$d_X$' contains the separation to the next Chandra X-ray source \citep{evans_chandra_2010}, if it is less than the positional uncertainties of the X-ray source.
Background galaxies are listed in Table~\ref{tbl:galaxies}.

\subsection{Cataclysmic variables}

   \begin{figure*}
   \centering
      \includegraphics[width=\textwidth]{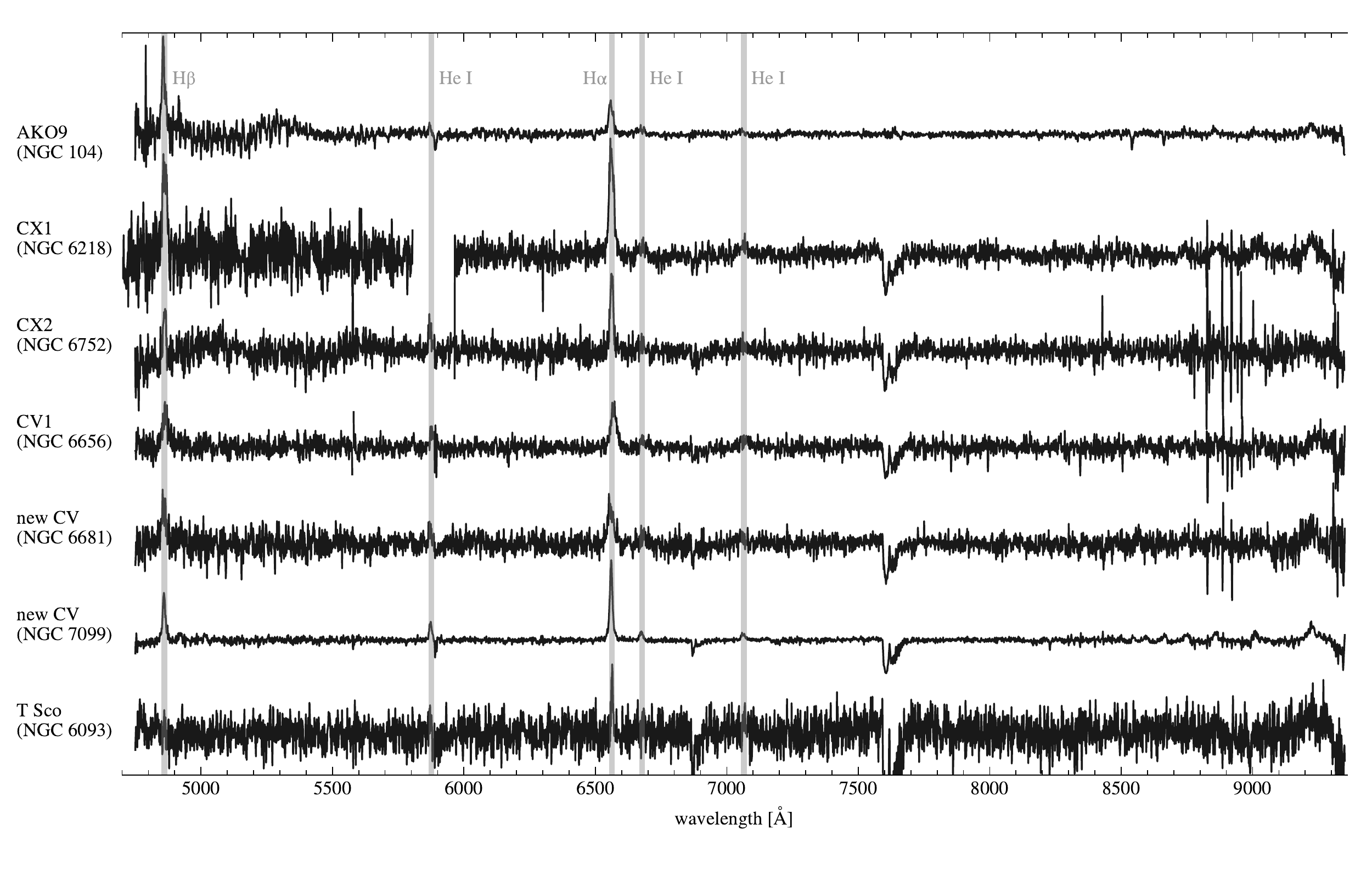}
      \caption{Normalised spectra of known cataclysmic variables AKO9, CX1, CX2, CV1, of the new CVs in \ngc{6681} and \ngc{7099}, and of the CV underlying the Nova T~Sco in 1860. 
      All spectra are detected by their broad Balmer emission lines, they also show \ion{He}{I} emission.
      The spectrum shown for AKO9 was created by combining several observed spectra.
              }
         \label{fig:cvs}
   \end{figure*}

As described above, cataclysmic variables (CVs) are binary systems consisting of a hot, compact white dwarf and a dwarf star in a close orbit. 
Only ten CVs have been confirmed by spectroscopy in the whole globular cluster system of the Milky Way \citep{knigge_cataclysmic_2012,webb_cv1_2013}.
Most CV candidates identified by photometry are not bright enough to be observed with MUSE and our relatively short exposure times. 
For example, \citet{sandoval_new_2018} lists $\text{R}_\text{625}$ magnitudes for 21 CV candidates in \ngc{104} of which only four are brighter than 20 mag.

\subsubsection{Known CVs and confirming CV candidates}
We find nine cataclysmic variables (CVs), of which seven are either previously spectroscopically confirmed CVs or candidates. 
The normalised MUSE spectra of several CVs including the previously unknown ones are shown in Fig.~\ref{fig:cvs}. 
Characteristic broad Balmer emission lines of {\Halpha} and {\Hbeta} are clearly visible in all spectra, as well as \ion{He}{i} emission.

One of them, CX1 in \ngc{6218} was detected as an X-ray source with optical counterpart and classified as a CV by \citet{lu_x-ray_2009} 
who consider it to be a member of the cluster based on its X-ray luminosity. 
We find that its optical counterpart is not the star marked in their finding chart but rather the star directly to the East with a F606W magnitude of 20.8.
With this new counterpart, we can confirm that CX1 is indeed a CV.

We do not see the characteristic broad Balmer emission lines for a CV in any spectrum of W56/X6 in \ngc{104} which was classified as a CV in \citet{heinke_deep_2005}.
However, the spectra show a \Halpha{} absorption line that is less deep than our spectral model predicts. 
Since this is not clearly a CV, we do not include it in our discussion in Section~\ref{sct:discussion_cv}.

\subsubsection{Nova T Scorpii}
In 1860, a classical nova in \ngc{6093} was observed by \citet{pogson_remarkable_1860}, Nova T Scorpii. 
Both \citet{shara_cataclysmic_1995} and \citet{dieball_far-ultraviolet_2010} looked for the underlying CV using near- and far-UV observations and they found a UV bright source at the right spatial position.
Using the finding charts in \citet{dieball_far-ultraviolet_2010}, we can identify their source~2129 with \acs{44184} \citep[$\text{F336W}-\text{F438W}=-0.1$, $\text{F438W}=18.5$,][]{piotto_hubble_2015,soto_hubble_2017}. 
This star was independently detected by our algorithm because of its broad \Halpha{} emission in several of its ten spectra observed with MUSE. 
A visual inspection shows that also \Hbeta{} and a weak \ion{He}{I} emission are present and variable. The \Halpha{} and \Hbeta{} lines seem to switch between emission and absorption.
However, as \acs{44184} is located on the lower RGB in the optical CMD, the CV has either a giant donor star or it is not resolved in the HST photometry but instead blended with a unrelated star.

\subsubsection{New cataclysmic variables}

Additionally to the seven known CVs, we find two more stars with very similar emission lines, indicating that these two stars are CVs as well.
One new CV is close to the centre of \ngc{7099} with a distance of {11\arcsec} and a F606W magnitude of 20.3 in the ACS catalogue (\acs{23423}).
Based on its position close to an X-ray source and its blue $U-V$ colour, \citet{lugger_chandra_2007} identified this star as a possible CV candidate (source C).
However, it is not included as a CV in the CVSGGC, which is why we list it as a new CV here.

The new CV in \ngc{6681} (\acs{19706}) has a distance of {27\arcsec} to the cluster centre and a F606W magnitude of 22.7.
The spectra of the new and known CVs are shown in Fig.~\ref{fig:cvs}.
Both new CVs are close ($0.13$ and $0.25\arcsec$, respectively) to a Chandra X-ray source listed in the November 2017 pre-release of the Chandra Source Catalog Release 2.0 \citep{evans_chandra_2010}.
Although \ngc{6681} was observed with HST in the UV \citep[see e.g.][]{massari_hubble_2013} and with the Chandra X-ray observatory \citep{pooley_chandra_2007,dieball_uncovering_2008}, 
no articles about CVs in this cluster have been published.

Are these CVs actually part of their respective cluster?
In general, we use the radial velocity and metallicity to determine if a star is a member of a globular cluster or a field star.
The standard spectral fit fails to determine reliable radial velocities or metallicities from the spectra of the new CVs.
We use a Gaussian fit to the {\Halpha} line to estimate the radial velocity for each CV, including the known CVs.  
The velocities differ from the cluster values by up to 300~km/s. 
This could be because the emission lines in some cases seem to have a more complex, that is non-gaussian, shape, or because of intrinsically high velocity variations due to the orbital motions,
or possibly because of eclipses of the accretion disk as observed for AKO9~\citep{knigge_far-ultraviolet_2003}.

We assume that CVs in GCs are spatially distributed in the same way as all other stars in a GC.
In the simulations of \citet{belloni_mocca-survey_2019}, 
CVs are either distributed more centrally or in the same way as main-sequence stars, depending on the relation time of the GCs.
With this property of CVs,
we can also use Bayes factors to decide between the two hypotheses $A \equiv$~``CV is a cluster member'' and $B \equiv$~``CV is not a cluster member''.
To calculate the factors we make use of the spatial distribution and membership probability 
derived from the observed radial velocity and metallicity of all the other stars observed with MUSE in the same field of view (see \citealt{kamann_stellar_2018} for details).
The distances of the new CVs to the respective cluster centre are about 1/6 of the half-light radius (\ngc{7099}) and 2/3 (\ngc{6681}) when the values from \citet{harris_catalog_1996} are used.
In the MUSE FOV of \ngc{7099} 96\,\% of all stars are cluster members; this leaves about $4\,\%$ non-members.
Of all member stars, 8\,\% are closer to the cluster centre than the CV we consider, while the remaining 92\,\% are farther away.
Thus, the likelihood of being a member star and at the same separation from the cluster centre or even closer is $p_A = 0.074.$ 
As for non-members, 10\,\% lie closer to the centre than the CV, and 90\,\% are farther away. This gives $p_B = 0.004.$
The Bayes factor is $p_A / p_B \approx 18$, which means that the positions of the CV provide evidence in favor of hypothesis $A$.
The same analysis for the CV that may be associated with \ngc{6681} gives a Bayes factor of $p_A / p_B \approx 13.$
Here, 92\,\% are member stars, of which two third lie closer to the centre than the CV. 
Of the 8\,\% non-members, 55\,\% lie closer to the centre.
According to the interpretation of \citet[p. 432]{jeffreys_theory_1998}, Bayes factors between $10$ and $10^{3/2}$ provide \textit{strong} evidence in favor 
of hypothesis $A$. 
In conclusion, the positions of the CVs and all the other stars in the MUSE FOV strongly suggest that both CVs are members of the respective cluster.

\subsection{Optical counterpart of M62-VLA1}

\label{section:black_holes}

   \begin{figure}
   \centering
      \includegraphics[width=0.5 \textwidth]{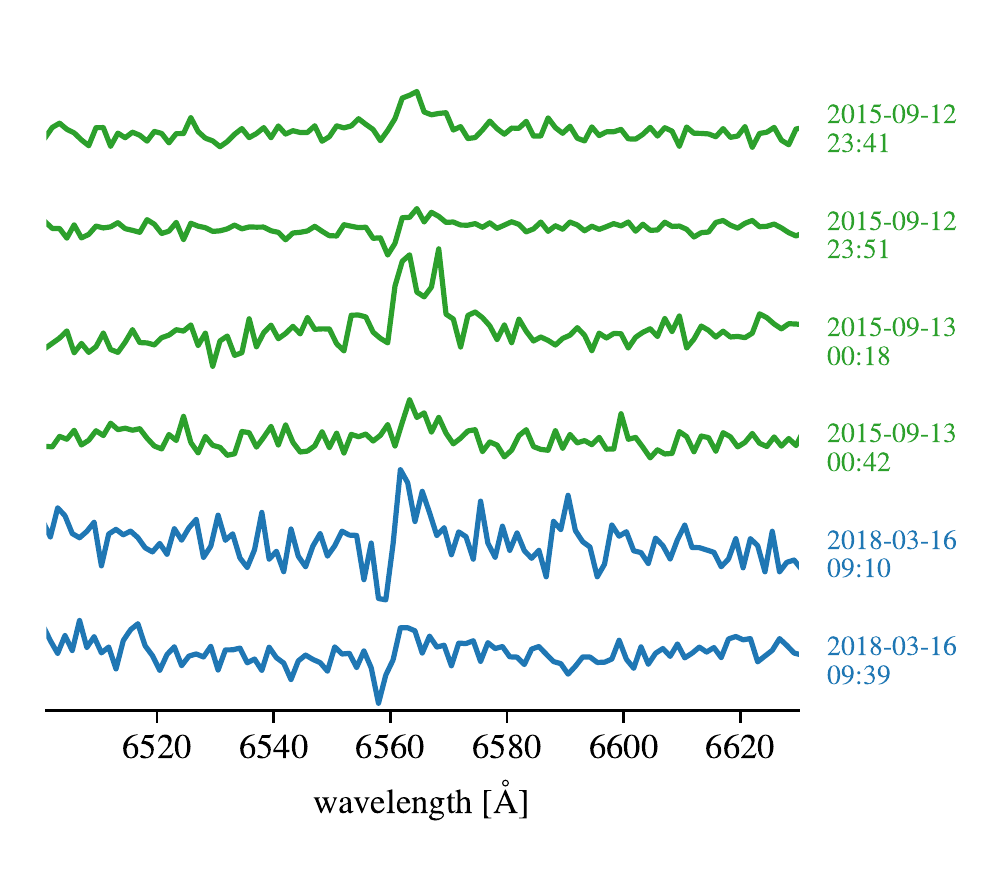}
      \caption{{\Halpha} region of multiple spectra of the likely companion of the black-hole candidate M62-VLA1. 
               This part of the spectrum is variable on the timescale of minutes.
              }
         \label{fig:bh_spectra}
   \end{figure}

Several stellar mass black holes or candidates are known in globular clusters, including three in \ngc{3201} \citep{giesers_detached_2018,giesers_stellar_2019}, 
M22-VLA1 and -VLA2 in \ngc{6656} \citep{strader_two_2012}, and M62-VLA1 in \ngc{6266} \citep{chomiuk_radio-selected_2013}.
The black-hole candidates in \ngc{3201} was discovered by \citet{giesers_detached_2018} and \citet{giesers_stellar_2019} using variations in the radial velocities of its visible companion observed with MUSE. 
These discoveries demonstrate that MUSE observations can be used to detect stellar exotica in GCs.

The black-hole candidate M62-VLA1 is close to the centre of \ngc{6266} and was discovered by \citet{strader_two_2012} using radio and X-ray observations.
It is likely to be part of a binary system with a star on the lower red-giant branch, which the authors identified in HST images very close to their radio source.
Our emission-line search found the optical counterpart of this black hole because of its {\Halpha} emission line. 
Both the position of this star and the counterpart reported in \citet{strader_two_2012} match, as well as the location in the colour-magnitude diagram.
This star has been observed several times with MUSE in 2015 and again in 2018 with varying signal-to-noise.
The spectra with the highest signal-to-noise show a {\Halpha} emission line which seem to vary between observations. 
These variations in the {\Halpha} region are shown in Fig.~\ref{fig:bh_spectra} where the shape of the emission line changes within tens of minutes.
We need more observations to determine reliable orbital parameters for this system 
similarly to the black holes and the 92 other binary systems in \ngc{3201} \citep{giesers_stellar_2019}.

\subsection{Red stragglers and sub-subgiants}
Red stragglers (RS) and sub-subgiants (SSG) are stars in globular clusters that occupy the region redwards of the red-giant branch or below the subgiant branch in the CMD. 
Since stellar evolution theory predicts these regions to be empty, their existence needs to be explained by more complicated formation theories \citep{geller_origin_2017,leiner_origin_2017,geller_origin_2017-1}.
Besides their location in the CMD, RS are X-ray and {\Halpha} emitters, photometrically variable and mostly radial-velocity binaries \citep{geller_origin_2017}.
As expected, several detected emission-line stars fall into the CMD region occupied by RS (four stars) and SSG (12 stars, see column `ID' in Table~\ref{tbl:elos}).
In particular, we detect the RS binary in \ngc{6254} discovered by \citet{shishkovsky_maveric_2018} which is also a source of radio and X-ray radiation.
A similar case is a star (\acs{40733}) in the RS region of \ngc{6541} which has a very broad and variable {\Halpha} emission, and it is close to an X-ray source ($0\farcs18$).
Of the eleven SSG that show \Halpha{} emission and are probable cluster members, eight are close to an X-ray source, eight show variations in their radial velocities, 
and seven SSG are both close to an X-ray source and have radial velocity variations.
All four RS with \Halpha{} emission are members of their respective cluster, three are close to an X-ray source and those three RS also show radial velocity variations.
We do not detect variability in the radial velocities of the fourth RS and it is not associated with an X-ray source.
These correlations fit the general characteristics of SSG and RS as described in \citet{geller_origin_2017}.
Orbital parameters for several SSG systems in \ngc{3201}, including the \Halpha{} emitters discovered here, are presented in \citet{giesers_stellar_2019}.

\subsection{Pulsating variables}
   
   \begin{figure}
   \centering
      \includegraphics[width=0.5 \textwidth]{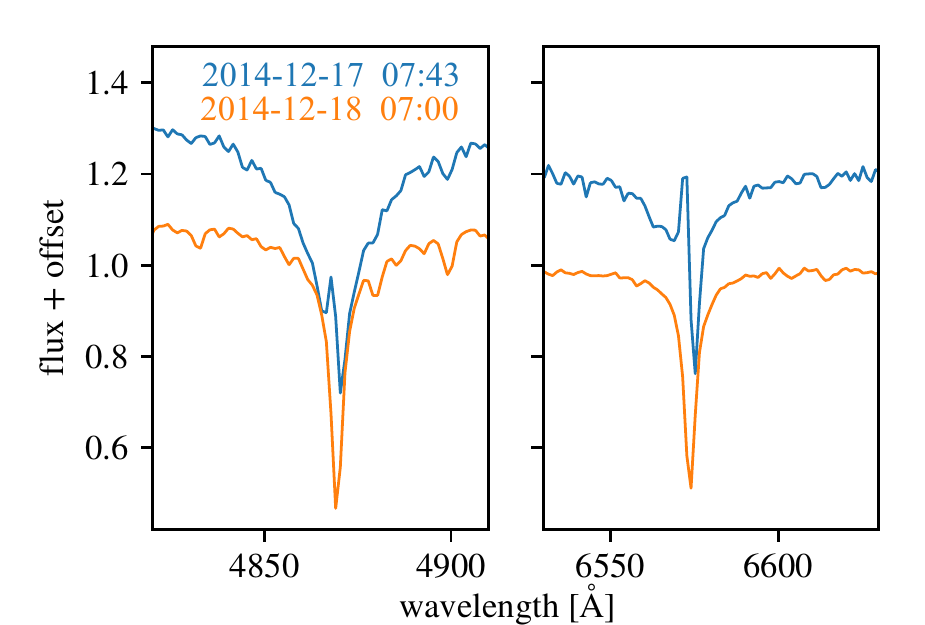}
      \caption{Two spectra of the RR Lyrae star NGC3201-V90 (\CVSGGC) observed approximately 24 hours apart show very different Balmer profiles.
               {\Halpha} and {\Hbeta} emission is only observed in one of the spectra, but not in the other one.
              }
         \label{fig:rrlyr_example}
   \end{figure}

About 40\,\% of all emission-line stars found in this survey are already known pulsating variable stars including
W~Viriginis variables, slow irregular variables, long-period variables, semiregular variables, and RR Lyrae variables.
Spectra of these stars show constant or variable \Halpha{} and sometimes \Hbeta{} emission fluxes.

Since the work of \citet{struve_peculiar_1947} it is known that spectra of RR Lyrae stars have a varying, weak emission component in several Hydrogen lines.
Figure~\ref{fig:rrlyr_example} shows two spectra of the RR Lyrae star V90 in \ngc{3201} as an example of variable emission lines.
The two spectra were observed roughly 24 hours apart and only the earlier one shows {\Halpha} and {\Hbeta} emission lines.
The flux profiles are very similar to the first or second apparition of the RR~Lyr variable X~Ari shown in \citet{gillet_emission_2014} in Fig.~1.
The variable star V13 and possibly also V190 in \ngc{6266} are currently classified as RR Lyrae stars in the literature.
However, their spectra show very strong \Halpha{} and \Hbeta{} emission lines, similar to those of W~Viriginis variables.

\subsection{Known nebulae}

   \begin{figure}
   \centering
      \includegraphics[width=0.5 \textwidth]{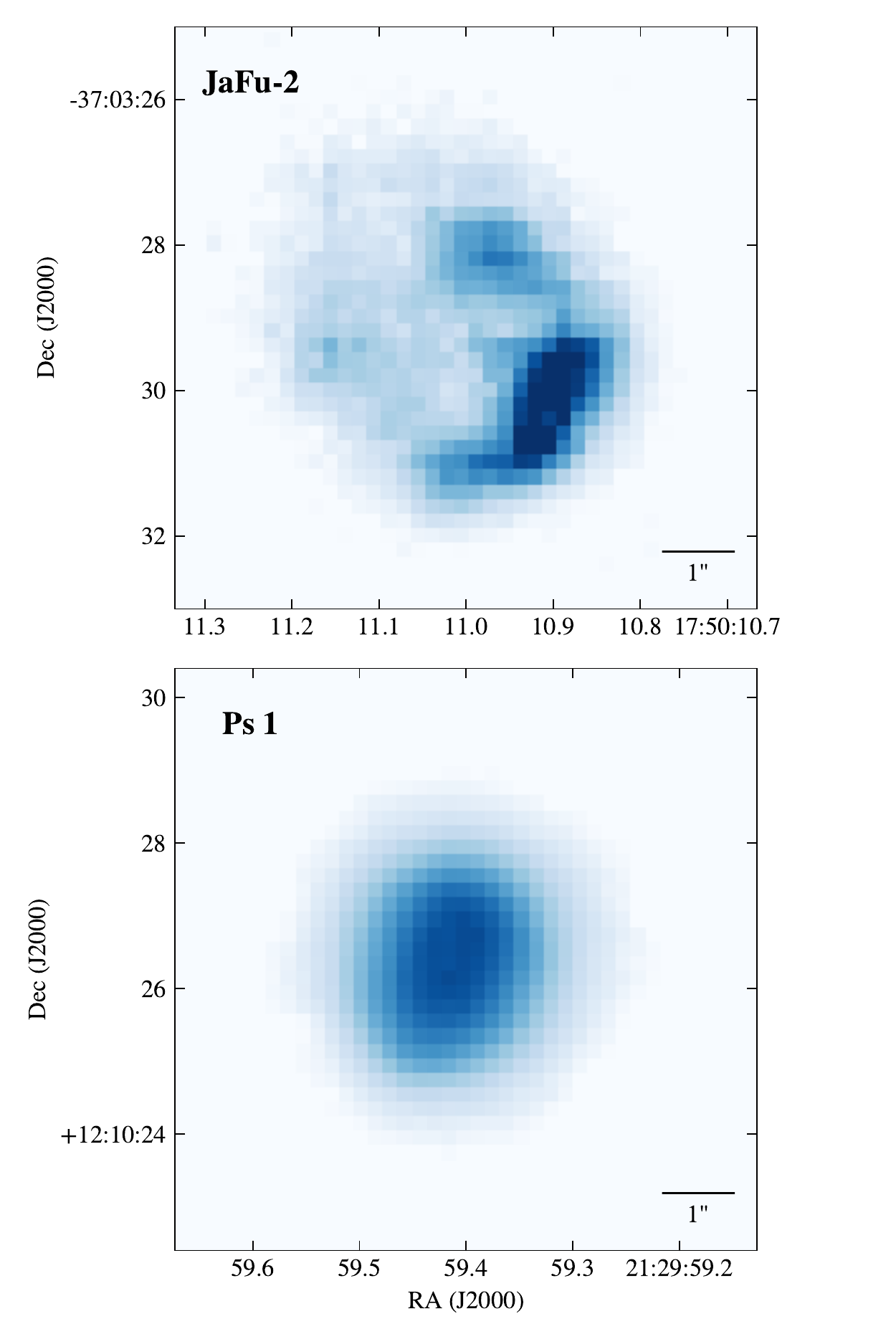}
      \caption{Flux maps of the two known planetary nebulae in our sample: JaFu-2 (top) in \ngc{6441} and Ps~1 (bottom) in \ngc{7078}. 
      Shown is the [\ion{O}{iii}]$\lambda5007$ flux after the stellar background has been subtracted.
              }
         \label{fig:known_pne}
   \end{figure}
   
      \begin{figure}
   \centering
      \includegraphics[width=0.5 \textwidth]{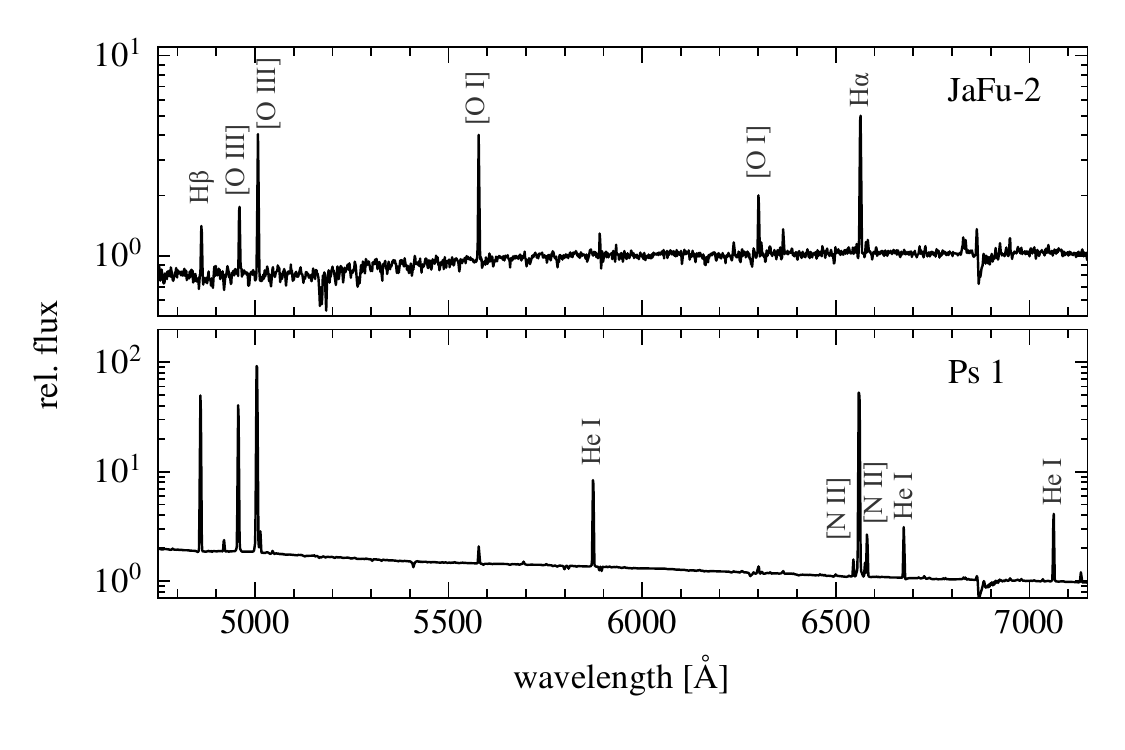}
      \caption{Spectra of JaFu-2 (top) and Ps~1 (bottom). Prominent emission lines of Hydrogen, \ion{He}{i}, {\NII}, [\ion{O}{i}] and [\ion{O}{iii}] are labelled.
              }
         \label{fig:known_pne_spectra}
   \end{figure}
   
There are four known planetary nebulae (PNe) in the whole globular cluster system of the Milky Way.
These are Ps~1 in \ngc{7078} \citep{pease_planetary_1928}, GJJC-1 in \ngc{6656} \citep{gillett_optical/infrared_1989}, JaFu-1 in Pal 6, and JaFu-2 in \ngc{6441} \citep[both][]{jacoby_planetary_1997}.

The most successful and also largest survey of PNe in globular clusters is the one from \citet{jacoby_planetary_1997}, who used the on-band/off-band technique at the [\ion{O}{iii}] line at 5007~{\AA} on 133 globular clusters.
With this survey, they doubled the number of known PNe in GCs from two to four.

Of the four globular clusters with known PNe, Pal 6 is not included in our sample of GCs. 
Although our survey covers \ngc{6656}, GJJC-1 is not inside the MUSE FOV because of its position inside the cluster.
This leaves Ps~1 and JaFu-2 for which we provide flux maps and spectra.
Figure~\ref{fig:known_pne} shows [\ion{O}{iii}] maps of the two PNe side-by-side, the PN spectra are shown in Fig.~\ref{fig:known_pne_spectra}.
The spectra were extracted with a relatively large circular aperture covering the whole nebula.
Although the nebulae are not HST point sources and, accordingly, we do not automatically extract a spectrum at their positions, 
the detection algorithm finds the nebular emission lines.
This is because their emission flux contaminates the otherwise purely stellar spectra of dozens of nearby stars 
as its high spatial variability is not accounted for in the extraction.

In addition to these two planetary nebulae, we detected a nova remnant in \ngc{6656} which is described in detail in \citet{gottgens_discovery_2019}.
However, we did not find any additional nebulae in our observations. We checked this null-result by stacking cubes from different observations to increase our sensitivity.
We used the on-band/off-band technique around \Halpha{} to search for extended emitting regions and did not find any nebula.

\subsection{Galaxies}
\label{sct:galaxies}

   \begin{figure*}
   \centering
      \includegraphics[width=\textwidth]{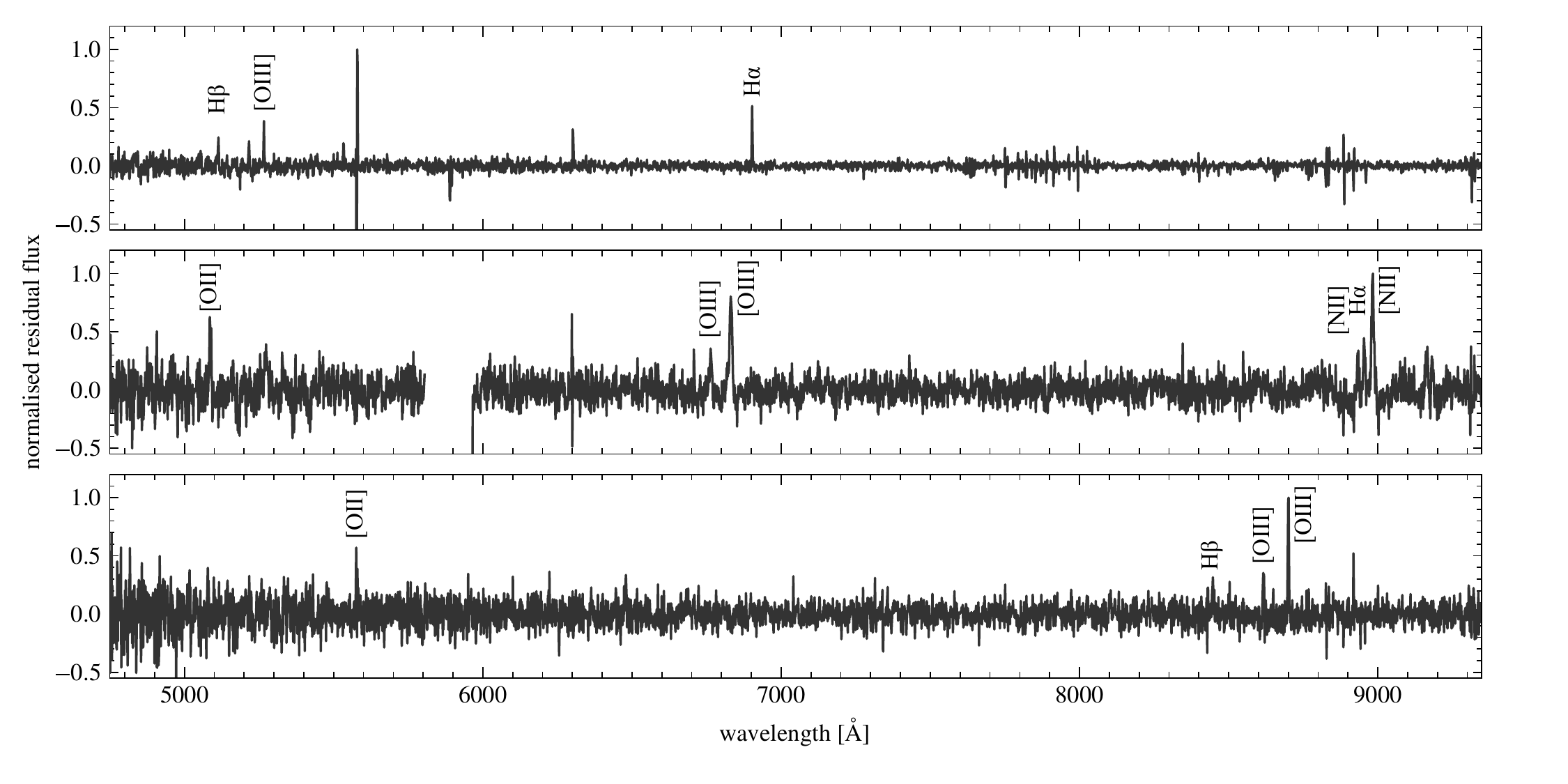}
      \caption{Normalised residuals of stellar spectra that contain extragalactic emission lines including from [\ion{O}{II}], [\ion{O}{III}], and \ion{H}{I}.
	      Redshifts determined from these lines are $z = 0.05, 0.36$ and $0.737$, respectively. The top and bottom panels show a starburst galaxy, while the middle panel shows an AGN.
              }
         \label{fig:galaxy_spectra}
   \end{figure*}

\begingroup
\renewcommand{\arraystretch}{1.2} 
\begin{table}
\tiny
\centering
\caption{Positions and redshifts of background galaxies with emission lines. The projected angular separation to the cluster centre is $d_C$.}
\begin{tabular}{lrrrr}
\hline
  Cluster &         RA [$\degr$] &       Dec [$\degr$] & $d_C$ [\arcsec] &      z \\
 \hline
  NGC 104 &    5.97861 & -72.10556 &           100.6 &  0.330 \\
  NGC 104 &    6.05634 & -72.05887 &            88.4 &  0.472 \\
 NGC 1851 &   78.52291 & -40.03877 &            31.5 &  0.309 \\
 NGC 1851 &   78.53530 & -40.03363 &            50.5 &  0.633 \\
 NGC 5139 &  201.66150 & -47.40166 &           293.4 &  0.144 \\
 NGC 5139 &  201.66193 & -47.39549 &           314.4 &  0.052 \\
 NGC 5904 &  229.61804 &   2.06133 &           102.0 &  0.414 \\
 NGC 6266 &  255.29045 & -30.10210 &            58.0 &  0.722 \\
 NGC 6388 &  264.07237 & -44.73578 &             1.8 &  0.420 \\
 NGC 6541 &  272.02490 & -43.70413 &            55.1 &  0.413 \\
 NGC 6681 &  280.79848 & -32.28750 &            21.9 &  0.305 \\
 NGC 6681 &  280.79929 & -32.29805 &            24.4 &  0.173 \\
 NGC 6752 &  287.66226 & -59.98960 &           100.4 &  0.364 \\
 NGC 6752 &  287.69412 & -59.96586 &            79.0 &  0.312 \\
 NGC 6752 &  287.69667 & -59.96068 &            93.5 &  0.246 \\
 NGC 6752 &  287.72543 & -60.01487 &           110.1 &  0.505 \\
 NGC 6752 &  287.77154 & -59.98476 &            98.0 &  0.105 \\
 NGC 7078 &  322.49858 &  12.17520 &            35.4 &  0.261 \\
 NGC 7078 &  322.50275 &  12.15647 &            51.0 &  0.670 \\
 NGC 7089 &  323.37114 &  -0.82728 &            34.0 &  0.737 \\
 NGC 7099 &  325.09295 & -23.17880 &             4.6 &  0.398 \\
\hline
\end{tabular}
\label{tbl:galaxies}
\end{table}
\endgroup

For each spectrum, we use the full list of emission-line candidates and their wavelengths to check if they correspond to a list of typical galactic emission lines, assuming they all have the same redshift.
Using this method, we find 21 background galaxies that contaminate spectra we have extracted at known stellar positions in observations of several GCs (see Table~\ref{tbl:galaxies}).
Since these spectra contain the stellar {\Halpha} absorption line, we conclude that these are indeed blended spectra of a star and a background galaxy. 
The spectra are identified by their prominent emission lines of Hydrogen and ionised Oxygen, as shown in Fig.~\ref{fig:galaxy_spectra} for three examples.
We detect emission lines corresponding to restframe wavelengths of 3727~\AA\ from [\ion{O}{II}], of 4959~\AA\ and 5007~\AA\ from [\ion{O}{III}], and {\Hbeta} and {\Halpha} emission.
Table~\ref{tbl:galaxies} lists the position, the redshift calculated from the emission lines, and the projected angular separation to the cluster centre for each galaxy.
Because of the low {\NII} to {\Halpha} flux, all galaxies fall into the region occupied by starburst galaxies in the BPT diagram \citep{baldwin_classification_1981,veilleux_spectral_1987}, 
except for one galaxy behind \ngc{6752} at $z=0.364$ which lies at the border of Seyfert galaxies and LINER.
Deeper photometric observations of the fields containing the new galaxies may identify the counterparts which
could be used to convert relative stellar proper motions into absolute proper motions, as done for \ngc{6681} using HST images \citep{massari_hubble_2013}.
These serendipitous discoveries resemble the one reported by \citet{bedin_hst_2019}, who found a dwarf spheroidal galaxy behind the globular cluster \ngc{6752} using HST photometry.
The fact that a galaxy lies very close to the core of the core-collapsed cluster \ngc{7099} shows our capability to look through the GCs.

\subsection{Unidentified sources}
\label{sct:unident_sources}

We find several emission-line stars that lie close to known X-ray sources.
As a reference, we use the November 2017 pre-release of the Chandra Source Catalog Release 2.0 \citep{evans_chandra_2010}, 
which includes positions and error ellipses (including astrometric uncertainties) for X-ray sources in all but two clusters in our survey (\ngc{6254} and \ngc{6624}).
Based on the source positions and the associated errors, we estimated that there is no physical relation between a star and an X-ray source if their distance is $> 1\arcsec.$
In general, multiple stars have a distance lower than $1\arcsec$ to an X-ray source which prohibits a unique identification of the optical counterpart.
However, because both emission-line sources and X-ray sources are rare objects in globular clusters, we indicate if an X-ray source is close in the Table~\ref{tbl:results}.

As indicated in Table~\ref{tbl:elos}, some stars show variable {\Halpha} emission wings or asymmetric absorption.
In the case of giant stars, these features could point to chromospheric activity or mass motions \citep[e.g.][]{cohen_mass_1976,cacciari_mass_2004,meszaros_mass_2008}.

\section{Discussion and conclusions}

\subsection{Completeness of the extraction of stellar sources}
\label{sct:discussion_cv}
There are two steps in the detection of emission-line sources in MUSE data that influence how many existing sources can be found:
the extraction completeness and the efficiency of matched filtering. We discuss only the first one here, the second one was described in Sect.~\ref{sct:efficiency}.

   \begin{figure}
   \centering
      \includegraphics[width=0.5\textwidth]{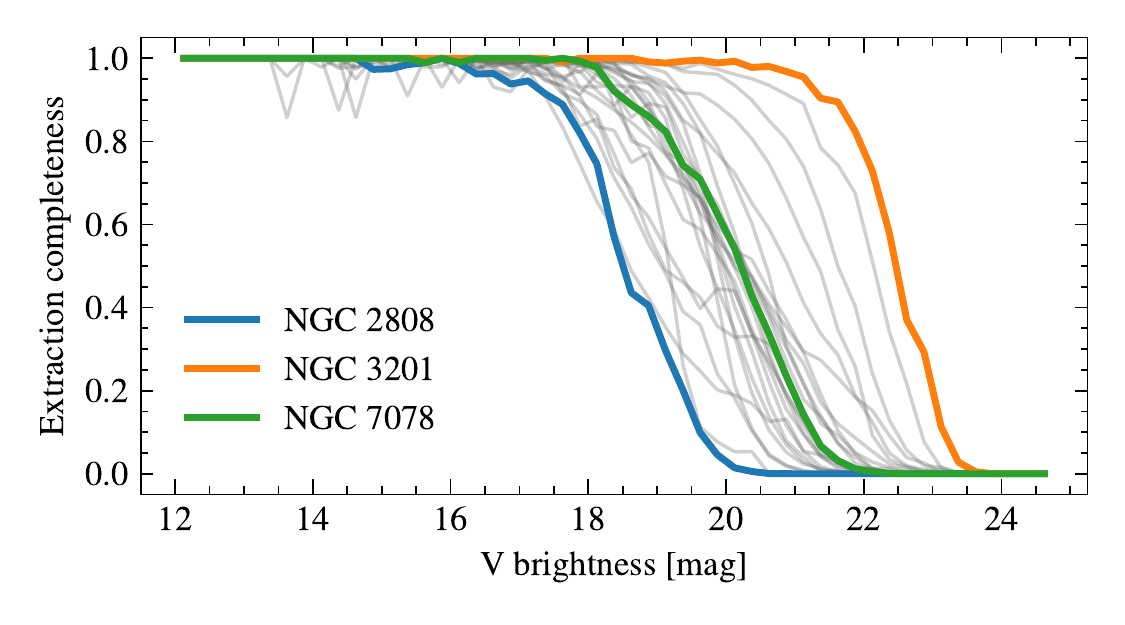}
      \caption{Spectral extraction completeness (SNR > 5) for all clusters with respect to the ACS catalogue as a function of brightness (filters are listed in Table~\ref{table:definition_vmag}). 
	       Each grey curve represents a single cluster, the curves for \ngc{2808}, \ngc{3201}, and \ngc{7078} are highlighted.
	       Since the completeness mainly depends on the stellar density, it also depends on the radial distance to the cluster centre (see Fig.~\ref{fig:completeness_ngc7078} in the appendix).
              }
         \label{fig:completeness_all_clusters}
   \end{figure}

The extraction completeness is the ratio of ACS catalogue sources for which a spectrum can be extracted from the MUSE datacube to all sources in the MUSE field of view.
Figure~\ref{fig:completeness_all_clusters} shows the dependence of our extraction completeness for all clusters taking all spectra with a SNR better than five into account.
An extraction completeness of 100\,\% does not mean that we have a spectrum of all stars in our FOV but only of those listed in the ACS catalogue.
This is an important distinction for low brightness stars in the central few arcseconds of core-collapsed clusters, such as \ngc{7078}.
Thus we expect that the completeness depends not only on the brightness of a star but also on its position relative to the cluster centre (see Fig.~{\ref{fig:completeness_ngc7078}}).

We estimate the extraction completeness for different magnitudes and for three regions: the whole FOV, the central 10\arcsec, the intermediate region from 10\arcsec to 60\arcsec and the remaining outer regions.
Figure~\ref{fig:completeness_ngc7078} in the Appendix shows how our extraction completeness depends on the brightness of the star and its position in the case of \ngc{7078}.
In general we see a completeness of close to 100~{\%} for bright stars throughout the cluster.
The crowded cluster centre hinders the extraction of faint sources and the completeness starts to drop below 50\,\%\ for magnitudes between 18 and 19 mag for most clusters.

\subsection{Do we find enough CVs?}
Massive globular clusters are expected to host about 200 CVs \citep{ivanova_formation_2006, knigge_cataclysmic_2012}. 
However, the cluster with the most CV candidates, as determined by UV and optical photometry and X-ray data, is \ngc{104} with 43 CVs \citep{sandoval_new_2018}.
In contrast, the number of spectroscopically confirmed CVs is much lower: 
only ten CVs have been confirmed by spectroscopy in the whole globular cluster system of the Milky Way \citep{knigge_cataclysmic_2012,webb_cv1_2013}.
We add seven CVs to this list, including two newly detected CVs.
CVs in globular clusters are hard to observe by spectroscopy because of the low brightness of the secondary component and crowding.
We expect to find dwarf novae (DNe), a subtype of CVs, because they have spectra with emission lines in quiescence 
\citep{clarke_evolution_1984,warner_dwarf_1995} and observations show that most CVs are DNe \citep{knigge_evolution_2011}. 
Is our number of CV detections consistent with the prediction? 
To answer this question, we use the average CV brightness distribution from MOCCA simulations of globular clusters \citep{belloni_mocca-survey_2016}. 
We did not consider effects of incomplete spatial coverage in our survey, because most CVs are expected to be located inside the half-mass radius of the respective cluster \citep[][but also see \citealt{belloni_mocca-survey_2019}]{belloni_mocca-survey_2016}.
For each cluster, we draw samples from this distribution and use our completeness function and a detection probability of 80\,\% (Sect.~\ref{sct:efficiency}) to estimate the number of CVs for which we should have extracted spectra (see Table~\ref{table:expected_cv} in the Appendix).
Since the clusters differ in their structural parameters and \citet{belloni_mocca-survey_2016} give a CV brightness distribution for an average globular cluster, 
the number of CVs for each individual cluster is probably not meaningful.
The total number of expected CV detections using the model of \citet{belloni_mocca-survey_2016} in our sample is $10 \pm 2$, which is consistent with our number of nine detected CVs.%

The step that restricts the overall completeness for CVs the most is the extraction completeness at magnitudes of 22 and below.
This could be improved with longer observations using the narrow-field mode (NFM) of MUSE, which will offer a much higher spatial sampling in a smaller FOV. 

\subsection{Exclusion of more PNe}
While the limiting fluxes given in Table~\ref{table:minimum_fluxes} are, if interpreted strictly, only valid for emission in stellar spectra, 
we can rule out any large diffuse source of {\Halpha} emission lines in our fields of view.
Any nebula or other diffuse source of emission would need to overlap with at least some stars of which we extracted spectra.
In the same way we easily detected JaFu-2, Ps~1, and the nova remnant in \ngc{6656} \citep{gottgens_discovery_2019}, these contaminated spectra would have been found.

There are still some possible but unlikely ways a hypothetical PNe could be hidden in a GC:
It could be very small so that it only contaminates a few stars, ideally of low brightness. In this case, we might not extract a spectrum for them.
The maximum size of this nebula cannot be large, considering the high density of sources for which we can extract spectra.
Another possibility is a nebula with very faint {\Halpha} emission that would have to be much fainter than the known ones because those were easily found with our detection method. 

However, it is still possible that more nebulae similar to the nova in \ngc{6656} lie outside the area covered by our survey.
In this case, it has to be in a region of relatively high stellar density where it can not be detected by photometric surveys. 

\section{Summary}

We analysed data from our MUSE survey of 26 Galactic globular clusters, looking for signs of emission-line objects.
Taking advantage of previous work on the same data, including data reduction, spectra extraction, and spectral analysis, we found {\nemstars} emission-line stars
and several non-stellar emission-line sources. 
By assuming a Gaussian emission-line shape and using matched filtering, we detect this shape in the residuals generated during a full spectrum fit to the observed spectra.
Since this generates many potentially interesting emission-line candidates, we use a threshold to select only the most promising candidates and check them visually.
We did not use external catalogues to search for known sources in our data, but we used them to validate and categorise our findings.
We found two new cataclysmic variables, many known pulsating variable stars, and several unidentified emission-line stars close to known X-ray sources.
The total number of CVs detected in this survey is consistent with numerical simulations when our spectral extraction completeness is taken into account.
In addition to stellar emission-line sources, we also found 20 previously unknown starburst galaxies and one AGN in the background with redshifts from 0.05 to 0.74.

\begin{acknowledgements}
FG, SK and SD acknowledge support from the German Research Foundation (DFG) through projects KA 4537/2-1 and DR 281/35-1. 
SK gratefully acknowledges funding from a European Research Council consolidator grant (ERC-CoG-646928- Multi-Pop).
PMW, SK, SD and BG also acknowledge support from the German Ministry for Education and Science (BMBF Verbundforschung) through projects MUSE-AO, grants 05A14BAC and 05A14MGA, and MUSE-NFM, grants 05A17MGA and 05A17BAA. 
Based on observations made with ESO Telescopes at the La Silla Paranal Observatory under programme IDs 094.D-0142, 095.D-0629, 096.D-0175, 097.D-0295, 098.D-0148, 099.D-0019, 0100.D-0161, 0101.D-0268, and 0102.D-0270.
Also based on observations made with the NASA/ESA Hubble Space Telescope, obtained from the data archive at the Space Telescope Science Institute. 
STScI is operated by the Association of Universities for Research in Astronomy, Inc. under NASA contract NAS 5-26555.
\end{acknowledgements}

\begingroup
\renewcommand{\arraystretch}{1.2} 
\begin{table}
\small
\caption{Choice of V filter for each cluster used in this paper.}
\label{table:definition_vmag}
\centering
\begin{tabular}{ll}
\hline 
\changed{ACS/WFC3} filter & Clusters \\
\hline

F625W & \ngc{6522} \\
F555W & \ngc{1904}, \ngc{6266}, \ngc{6293} \\
F606W & \ngc{104}, \ngc{1851}, \ngc{2808}, \ngc{3201} \\
& \ngc{362}, \ngc{5139}, \ngc{5286}, \ngc{5904}, \\
& \ngc{6093}, \ngc{6121}, \ngc{6218}, \ngc{6254}, \\
& \ngc{6388}, \ngc{6441}, \ngc{6541}, \ngc{6624} \\
& \ngc{6656}, \ngc{6681}, \ngc{6752}, \ngc{7078} \\
& \ngc{7089}, \ngc{7099} \\

\hline
\end{tabular}
\end{table}
\endgroup

\bibliographystyle{aa}
\bibliography{elo_paper2}

\begin{appendix}
\section{Completeness}
   \begin{figure}
   \centering
      \includegraphics[width=0.5\textwidth]{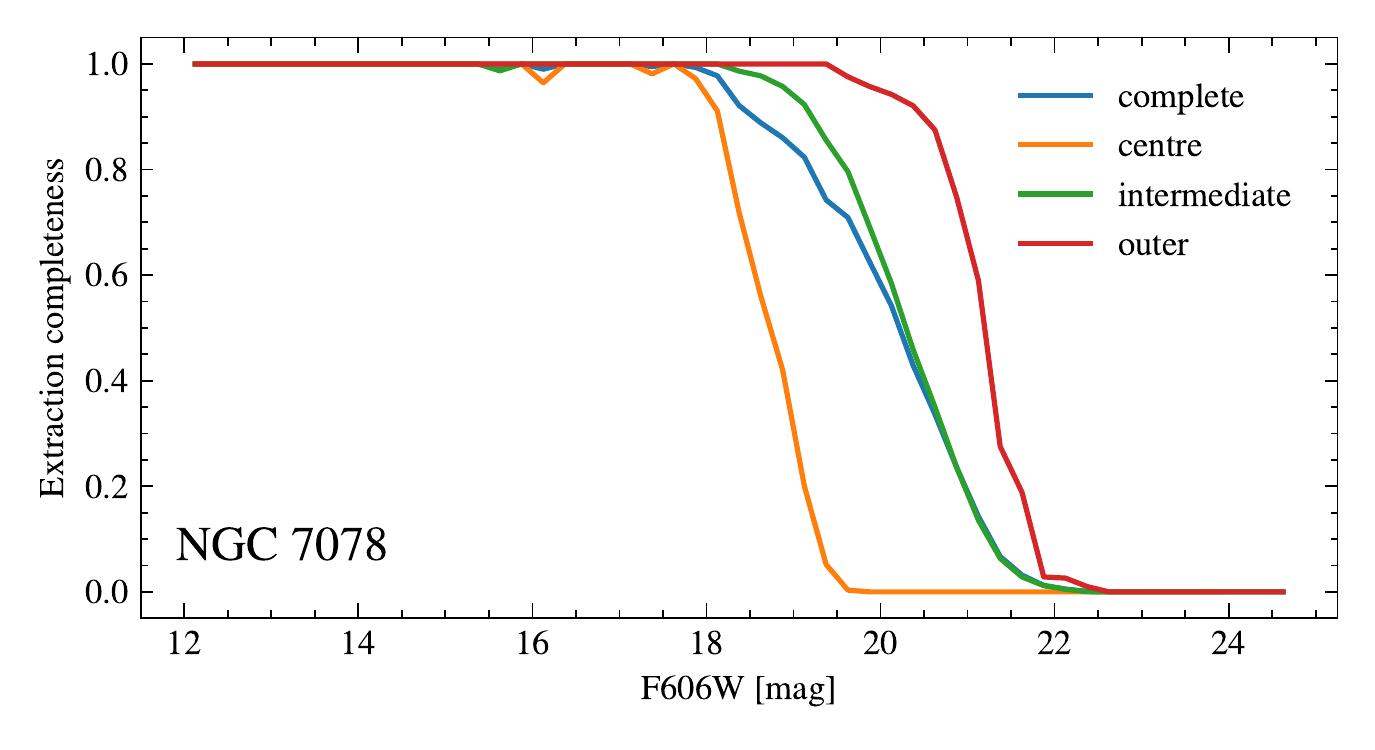}
      \caption{Spectral extraction completeness in \ngc{7078} relative to the ACS catalogue. For this cluster, we reach a completeness of 50\,\% at a magnitude of 17, 20, and 21.5
      for the central region (innermost 10\arcsec), an intermediate region (between 10\arcsec and 60\arcsec), and the outer regions (outside 60\arcsec), respectively.
              }
         \label{fig:completeness_ngc7078}
   \end{figure}
\begingroup
\renewcommand{\arraystretch}{1.2} 
\begin{table}
 \caption{\changed{Number of expected CVs in each cluster using the brightness distribution from \citet{belloni_mocca-survey_2016}. 
 Columns `lower' and `upper' give the $1\sigma$ interval around the median value listed in column `median'. 
 The column $N_{\rm found}$ refers to the number of spectroscopically detected CVs in our survey.}}
\label{table:expected_cv}
\centering

\begin{tabular}{lrrrl}
\hline
    Cluster & lower & median & upper & $N_{\rm found}$ \\
\hline
 \ngc{3201} &      1.5 &      2.9 &      4.8 &                 \\
 \ngc{6218} &      0.9 &      1.7 &      3.3 &               1 \\
 \ngc{6254} &      0.8 &      1.6 &      2.9 &                 \\
 \ngc{6656} &      0.5 &      1.3 &      2.1 &               1 \\
 \ngc{6752} &      0.4 &      0.9 &      1.4 &               1 \\
 \ngc{6121} &      0.1 &      0.3 &      0.9 &                 \\
  \ngc{104} &      0.0 &      0.1 &      0.3 &               1 \\
 \ngc{5904} &      0.0 &      0.1 &      0.3 &                 \\
 \ngc{6681} &      0.0 &      0.1 &      0.3 &               1 \\
 \ngc{7099} &      0.0 &      0.1 &      0.6 &               1 \\
 \ngc{6266} &      0.0 &      0.0 &      0.2 &                 \\
 \ngc{6624} &      0.0 &      0.0 &      0.2 &                 \\
 \ngc{5139} &      0.0 &      0.0 &      0.2 &               1 \\
 \ngc{6541} &      0.0 &      0.0 &      0.1 &                 \\
 \ngc{5286} &      0.0 &      0.0 &      0.0 &                 \\
 \ngc{6388} &      0.0 &      0.0 &      0.0 &                 \\
 \ngc{6441} &      0.0 &      0.0 &      0.0 &                 \\
 \ngc{7078} &      0.0 &      0.0 &      0.0 &                 \\
 \ngc{6093} &      0.0 &      0.0 &      0.0 &               2 \\
 \ngc{1851} &      0.0 &      0.0 &      0.0 &                 \\
 \ngc{6293} &      0.0 &      0.0 &      0.0 &                 \\
 \ngc{6522} &      0.0 &      0.0 &      0.0 &                 \\
  \ngc{362} &      0.0 &      0.0 &      0.0 &                 \\
 \ngc{2808} &      0.0 &      0.0 &      0.0 &                 \\
 \ngc{1904} &      0.0 &      0.0 &      0.0 &                 \\
 \ngc{7089} &      0.0 &      0.0 &      0.0 &                 \\
 total     &       8.1 &     10.0 &     12.5 &               9 \\
\hline
\end{tabular}
\end{table}

\endgroup

\clearpage
\onecolumn

\begin{landscape}
\small
\setlength{\tabcolsep}{0.5\tabcolsep}

\begin{longtable}{lrr lrr llrl ll}
\caption{\label{tbl:elos} Stars with spectra containing significant emission lines.} \\

\hline
Cluster & RA & Dec & ACS ID & $d_C$ [\arcsec] & V [mag] & mem.? & vrad? & $d_X$ [\arcsec] & Comment & Ident. & Ref. \\
\hline
\endfirsthead

\caption{cont.} \\
\hline
Cluster & RA & Dec & ACS ID & $d_C$ [\arcsec] & V [mag] & mem.? & vrad? & $d_X$ [\arcsec] & Comment & Ident. & Ref. \\
\hline
\endhead

\hline
\endfoot

\endlastfoot
NGC 104 & 0\nfh  24\nfm  2\fs17 & -72\degr  5\arcmin  42\farcs06 & 22645 & 52.0 & 17.1 & yes & ? & 0.12 & variable, filled \Halpha{} absorption & W56/X6 (CV) & 1 \\
NGC 104 & 0\nfh  24\nfm  4\fs21 & -72\degr  4\arcmin  43\farcs52 & 103550 & 11.3 & 16.9 & yes &  & 0.17 & variable \Halpha{} emission & (SSG) &  \\
NGC 104 & 0\nfh  24\nfm  4\fs91 & -72\degr  4\arcmin  55\farcs53 & 59082 & 4.6 & 16.6 & ? &  & 0.14 & broad \Halpha{}, \Hbeta{} and He emission & AKO9 (CV) & 2 \\
NGC 104 & 0\nfh  24\nfm  5\fs65 & -72\degr  5\arcmin  4\farcs21 & 58167 & 11.6 & 15.9 & yes &  & 0.15 & noisy, filled \Halpha{} absorption, contaminated? & W31 (MSP?) & 1 \\
NGC 104 & 0\nfh  24\nfm  6\fs61 & -72\degr  4\arcmin  50\farcs88 & 99987 & 4.7 & 15.5 & yes & yes & 0.36 & variable \Halpha{} emission & (SSG) &  \\
NGC 104 & 0\nfh  24\nfm  8\fs55 & -72\degr  5\arcmin  35\farcs16 & 53591 & 44.6 & 17.0 & yes & yes & 0.19 & noisy, variable, filled \Halpha{} absorption & WF4-V04 (EA) & 2 \\
NGC 104 & 0\nfh  24\nfm  8\fs58 & -72\degr  3\arcmin  54\farcs83 & 136338 & 59.3 & 11.0 & yes & no &  & variable \Halpha{}, \Hbeta{} emission & V1 (SR) & 2 \\
NGC 104 & 0\nfh  24\nfm  13\fs76 & -72\degr  3\arcmin  34\farcs56 & 132746 & 86.5 & 17.2 & yes & yes & 0.20 & variable, filled \Halpha{} absorption & (SSG) &  \\
NGC 104 & 0\nfh  24\nfm  15\fs16 & -72\degr  4\arcmin  43\farcs64 & 89607 & 44.7 & 17.0 & yes & yes & 0.14 & variable \Halpha{} emission & (RS) & 3 \\
NGC 104 & 0\nfh  24\nfm  17\fs23 & -72\degr  3\arcmin  2\farcs89 & 155735 & 122.0 & 17.0 & yes & yes & 0.10 & filled \Halpha{} absorption & (SSG?) &  \\
NGC 362 & 1\nfh  3\nfm  12\fs77 & -70\degr  51\arcmin  17\farcs81 & 44740 & 23.4 & 17.8 & yes & yes & 0.09 & variable \Halpha{} emission & (SSG) &  \\
NGC 362 & 1\nfh  3\nfm  14\fs42 & -70\degr  51\arcmin  1\farcs63 & 42163 & 6.1 & 16.0 & yes & yes &  & filled \Halpha{} absorption &  &  \\
NGC 362 & 1\nfh  3\nfm  14\fs89 & -70\degr  51\arcmin  28\farcs46 & 41200 & 33.0 & 16.3 & yes & yes & 0.07 & variable, filled \Halpha{} absorption & (RS) &  \\
NGC 362 & 1\nfh  3\nfm  15\fs08 & -70\degr  50\arcmin  32\farcs33 & 83177 & 23.6 & 11.8 & yes & yes &  & variable \Halpha{}, \Hbeta{} emission & V16 (Mira Ceti) & 2 \\
NGC 362 & 1\nfh  3\nfm  16\fs46 & -70\degr  50\arcmin  55\farcs96 & 39120 & 10.9 & 17.2 & yes & yes & 0.06 & variable \Halpha{} emission &  &  \\
NGC 362 & 1\nfh  3\nfm  17\fs38 & -70\degr  50\arcmin  37\farcs67 & 79642 & 23.6 & 15.2 & yes & yes &  & variable, filled \Halpha{} absorption & Sz54 (RRLyr) & 2 \\
NGC 362 & 1\nfh  3\nfm  22\fs11 & -70\degr  50\arcmin  44\farcs76 & 73827 & 40.2 & 16.2 & yes & yes & 0.09 & variable \Halpha{} emission & V24 (EA) & 2 \\
NGC 1851 & 5\nfh  14\nfm  3\fs70 & -40\degr  3\arcmin  5\farcs46 & 58086 & 39.4 & 16.1 & yes & yes &  & variable \Halpha{} absorption & V27 (RRLyr) & 2 \\
NGC 1851 & 5\nfh  14\nfm  6\fs43 & -40\degr  2\arcmin  48\farcs66 & 96334 & 3.8 & 13.0 & yes & ? &  & variable \Halpha{} absorption &  &  \\
NGC 1851 & 5\nfh  14\nfm  8\fs48 & -40\degr  3\arcmin  0\farcs98 & 41019 & 24.0 & 19.6 & yes &  &  & variable \Halpha{} emission, contaminated? &  &  \\
NGC 1851 & 5\nfh  14\nfm  8\fs54 & -40\degr  3\arcmin  2\farcs86 & 40603 & 25.6 & 19.0 & yes &  &  & noisy, filled \Halpha{} absorption &  &  \\
NGC 1851 & 5\nfh  14\nfm  8\fs55 & -40\degr  2\arcmin  16\farcs98 & 87919 & 36.9 & 15.4 & yes & yes &  & variable, filled \Halpha{} absorption & V4 (RRLyr) & 2 \\
NGC 1851 & 5\nfh  14\nfm  9\fs09 & -40\degr  2\arcmin  1\farcs21 & 86191 & 53.6 & 16.2 & yes & yes &  & variable, filled \Halpha{} and \Hbeta{} absorption & V15 (RRLyr) & 2 \\
NGC 1851 & 5\nfh  14\nfm  9\fs94 & -40\degr  2\arcmin  14\farcs94 & 83855 & 49.0 & 18.9 & yes & yes &  & variable \Halpha{} emission &  &  \\
NGC 1904 & 5\nfh  24\nfm  10\fs19 & -24\degr  31\arcmin  2\farcs45 & 250125 & 29.2 & 16.5 & yes & yes &  & variable \Halpha{}, \Hbeta{} absorption & V5 (RRLyr) & 2 \\
NGC 1904 & 5\nfh  24\nfm  11\fs48 & -24\degr  31\arcmin  24\farcs55 & 151206 & 7.0 & 14.6 & yes &  &  & variable \Halpha{} emission, contaminated? &  &  \\
NGC 1904 & 5\nfh  24\nfm  11\fs49 & -24\degr  31\arcmin  37\farcs09 & 4 & 9.8 & 12.6 & yes &  &  & \Halpha{} and \Hbeta{} emission & V8 (SR) & 2 \\
NGC 1904 & 5\nfh  24\nfm  12\fs63 & -24\degr  31\arcmin  40\farcs94 & 80378 & 24.3 & 13.5 & yes & yes &  & variable, filled \Halpha{} absorption & V7 (W Vir) & 2 \\
NGC 2808 & 9\nfh  12\nfm  2\fs35 & -64\degr  52\arcmin  4\farcs42 & 106976 & 16.5 & 17.2 & yes & yes &  & filled \Halpha{} absorption?, between RGB and HB &  &  \\
NGC 2808 & 9\nfh  12\nfm  3\fs51 & -64\degr  51\arcmin  43\farcs99 & 201163 & 5.3 & 12.5 & yes & yes &  & variable \Halpha{} emission &  &  \\
NGC 2808 & 9\nfh  12\nfm  5\fs53 & -64\degr  51\arcmin  31\farcs01 & 194298 & 23.4 & 14.7 & yes & yes &  & variable \Halpha{} emission & V51 (BL Her) & 4 \\
NGC 2808 & 9\nfh  12\nfm  6\fs72 & -64\degr  52\arcmin  40\farcs30 & 26920 & 56.6 & 12.2 & yes & no &  & variable \Halpha{} emission & V11 (SR) & 2 \\
NGC 3201 & 10\nfh  17\nfm  32\fs62 & -46\degr  25\arcmin  32\farcs58 & 6558 & 64.4 & 14.7 & yes & yes &  & variable, filled \Halpha{} absorption & V23 (RRLyr) & 2 \\
NGC 3201 & 10\nfh  17\nfm  33\fs14 & -46\degr  25\arcmin  7\farcs42 & 14749 & 44.2 & 17.0 & yes & yes &  & filled \Halpha{} absorption & V141 (EA) & 2 \\
NGC 3201 & 10\nfh  17\nfm  34\fs89 & -46\degr  24\arcmin  33\farcs15 & 23979 & 23.1 & 14.4 & yes & yes &  & variable \Halpha{}, \Hbeta{} absorption & V90 (RRLyr) & 2 \\
NGC 3201 & 10\nfh  17\nfm  35\fs59 & -46\degr  24\arcmin  50\farcs51 & 13438 & 13.9 & 17.2 & yes & yes & 0.14 & filled \Halpha{} absorption & (SSG) &  \\
NGC 3201 & 10\nfh  17\nfm  36\fs01 & -46\degr  25\arcmin  12\farcs50 & 13108 & 28.8 & 17.7 & yes & yes &  & filled \Halpha{} absorption & VN2 (EW) & 5 \\
NGC 3201 & 10\nfh  17\nfm  36\fs11 & -46\degr  25\arcmin  36\farcs03 & 5081 & 51.6 & 14.9 & yes & yes &  & variable, filled \Halpha{} absorption & V77 (RRLyr) & 2 \\
NGC 3201 & 10\nfh  17\nfm  36\fs14 & -46\degr  24\arcmin  28\farcs08 & 23379 & 18.2 & 14.9 & yes & yes &  & variable, filled \Halpha{}, \Hbeta{} absorption & V100 (RRLyr) & 2 \\
NGC 3201 & 10\nfh  17\nfm  37\fs58 & -46\degr  23\arcmin  52\farcs24 & 22692 & 53.2 & 17.3 & yes & yes & 0.06 & variable \Halpha{} emission & (SSG) &  \\
NGC 3201 & 10\nfh  17\nfm  38\fs56 & -46\degr  24\arcmin  27\farcs12 & 22186 & 25.3 & 18.7 & no & ? & 0.23 & filled \Halpha{} absorption &  &  \\
NGC 3201 & 10\nfh  17\nfm  39\fs11 & -46\degr  24\arcmin  30\farcs11 & 21907 & 27.9 & 18.4 & yes & yes &  & filled \Halpha{} absorption, two spectral components & V135 (EW?) & 2 \\
NGC 3201 & 10\nfh  17\nfm  39\fs14 & -46\degr  24\arcmin  11\farcs03 & 21922 & 41.5 & 18.4 & yes & yes &  & variable \Halpha{} emission &  &  \\
NGC 3201 & 10\nfh  17\nfm  39\fs25 & -46\degr  25\arcmin  11\farcs89 & 11405 & 36.9 & 17.3 & yes & yes &  & filled \Halpha{} absorption & (SSG) &  \\
NGC 5139 & 13\nfh  26\nfm  32\fs57 & -47\degr  30\arcmin  3\farcs77 & 2036152 & 167.5 & 19.7 \tablefootmark{a} & yes & yes &  & filled \Halpha{} absorption &  &  \\
NGC 5139 & 13\nfh  26\nfm  42\fs55 & -47\degr  24\arcmin  21\farcs87 & 1000193 & 268.9 & 14.2 \tablefootmark{a} & yes & yes &  & variable filled \Halpha{} absorption & V51 (RRLyr) & 2 \\
NGC 5139 & 13\nfh  26\nfm  43\fs30 & -47\degr  28\arcmin  16\farcs33 & 210583 & 50.0 & 18.7 & yes &  & 0.22 & variable \Halpha{} emission &  &  \\
NGC 5139 & 13\nfh  26\nfm  43\fs61 & -47\degr  29\arcmin  38\farcs37 & 66994 & 63.6 & 11.0 & yes & yes &  & variable \Halpha{} emission & LW10 (LPV) & 6 \\
NGC 5139 & 13\nfh  26\nfm  44\fs32 & -47\degr  29\arcmin  5\farcs23 & 113810 & 35.0 & 11.3 & yes & yes &  & variable \Halpha{} emission & LW11 (LPV) & 6 \\
NGC 5139 & 13\nfh  26\nfm  46\fs36 & -47\degr  29\arcmin  30\farcs51 & 65282 & 44.9 & 13.3 & ? & no &  & variable \Halpha{}, \Hbeta{} emission & V42 (SR) & 2 \\
NGC 5139 & 13\nfh  26\nfm  47\fs72 & -47\degr  29\arcmin  29\farcs00 & 64316 & 42.8 & 11.1 & yes & ? &  & variable \Halpha{} emission & V152 (SR) & 2 \\
NGC 5139 & 13\nfh  26\nfm  53\fs51 & -47\degr  29\arcmin  0\farcs24 & 125876 & 65.1 & 20.0 & yes &  & 0.15 & broad \Halpha{} emission & 13a (CV) & 7 \\
NGC 5139 & 13\nfh  26\nfm  55\fs04 & -47\degr  28\arcmin  44\farcs63 & 149184 & 79.1 & 18.9 & yes & yes &  & filled \Halpha{} absorption &  &  \\
NGC 5286 & 13\nfh  46\nfm  24\fs86 & -51\degr  22\arcmin  56\farcs89 & 80568 & 34.8 & 13.0 & yes &  &  & broad \Halpha{} emission, but close to detector edge & V59 (SR?) & 2 \\
NGC 5286 & 13\nfh  46\nfm  26\fs58 & -51\degr  22\arcmin  10\farcs29 & 142141 & 17.1 & 19.3 & yes &  &  & filled \Halpha{} absorption & (SSG) &  \\
NGC 5286 & 13\nfh  46\nfm  27\fs01 & -51\degr  22\arcmin  29\farcs94 & 71539 & 3.2 & 13.1 & yes & yes &  & variable \Halpha{}, \Hbeta{} emission & V74 (L) & 2 \\
NGC 5286 & 13\nfh  46\nfm  27\fs08 & -51\degr  22\arcmin  30\farcs68 & 70971 & 4.2 & 19.8 & yes &  &  & noisy, filled \Halpha{} absorption &  &  \\
NGC 5286 & 13\nfh  46\nfm  27\fs73 & -51\degr  22\arcmin  17\farcs34 & 137375 & 13.2 & 18.6 & yes & ? &  & variable filled \Halpha{} absorption & (RS) &  \\
NGC 5286 & 13\nfh  46\nfm  28\fs99 & -51\degr  22\arcmin  51\farcs69 & 63415 & 31.9 & 13.1 & yes & yes &  & variable \Halpha{} emission & V60 (SR?) & 2 \\
NGC 5286 & 13\nfh  46\nfm  30\fs24 & -51\degr  22\arcmin  7\farcs37 & 127924 & 37.8 & 13.5 & yes &  &  & \Halpha{} emission, but close to detector edge &  &  \\
NGC 5904 & 15\nfh  18\nfm  27\fs42 & 2\degr  3\arcmin  24\farcs67 & 24106 & 122.9 & 14.5 & yes &  &  & variable, filled \Halpha{} absorption & V83 (RRLyr) & 2 \\
NGC 5904 & 15\nfh  18\nfm  32\fs15 & 2\degr  5\arcmin  2\farcs00 & 82524 & 19.0 & 16.1 & yes & yes & 0.15 & variable, filled \Halpha{} absorption &  &  \\
NGC 5904 & 15\nfh  18\nfm  32\fs85 & 2\degr  5\arcmin  41\farcs49 & 80224 & 50.1 & 15.2 & yes &  &  & variable, filled \Halpha{} absorption & V5 (RRLyr) & 2 \\
NGC 5904 & 15\nfh  18\nfm  33\fs74 & 2\degr  4\arcmin  6\farcs55 & 41637 & 45.8 & 21.2 & yes & yes &  & noisy filled \Halpha{} absorption &  &  \\
NGC 5904 & 15\nfh  18\nfm  34\fs29 & 2\degr  5\arcmin  1\farcs79 & 75350 & 19.1 & 16.4 & yes & yes & 0.15 & variable, filled \Halpha{} absorption &  &  \\
NGC 5904 & 15\nfh  18\nfm  36\fs12 & 2\degr  4\arcmin  14\farcs19 & 34416 & 57.4 & 19.3 & yes &  &  & variable \Halpha{} emission & (BSS) &  \\
NGC 5904 & 15\nfh  18\nfm  36\fs14 & 2\degr  4\arcmin  16\farcs30 & 34419 & 56.4 & 11.8 & yes &  &  & variable \Halpha{} emission & V84 (RV Tau) & 2 \\
NGC 6093 & 16\nfh  17\nfm  1\fs60 & -22\degr  58\arcmin  29\farcs20 & 96954 & 12.1 & 19.6 & yes &  & 0.22 & \Halpha{} emission & CX3 (CV) & 8 \\
NGC 6093 & 16\nfh  17\nfm  2\fs18 & -22\degr  58\arcmin  46\farcs10 & 47840 & 12.6 & 18.6 & no &  &  & noisy, variable, filled \Halpha{} absorption &  &  \\
NGC 6093 & 16\nfh  17\nfm  2\fs56 & -22\degr  58\arcmin  46\farcs31 & 45658 & 12.6 & 17.5 & yes & yes & 0.21 & variable, filled \Halpha{} absorption & CX12 (?) & 8 \\
NGC 6093 & 16\nfh  17\nfm  2\fs82 & -22\degr  58\arcmin  33\farcs88 & 44184 & 5.7 & 17.3 & yes & ? & 0.18 & variable \Halpha{} emission & T Sco (CV) & 9 \\
NGC 6121 & 16\nfh  23\nfm  31\fs46 & -26\degr  30\arcmin  57\farcs83 & 9918 & 61.2 & 16.6 & ? &  & 0.19 & \Halpha{} emission & 107/N33 (EC) & 2 \\
NGC 6121 & 16\nfh  23\nfm  32\fs40 & -26\degr  30\arcmin  45\farcs47 & 9674 & 60.5 & 17.4 & ? &  & 0.15 & filled \Halpha{} absorption & 108/N36 (?) & 2 \\
NGC 6121 & 16\nfh  23\nfm  35\fs04 & -26\degr  31\arcmin  19\farcs03 & 8911 & 13.9 & 16.1 & yes &  & 0.19 & filled \Halpha{} absorption & CX10 (AB) & 10 \\
NGC 6121 & 16\nfh  23\nfm  35\fs07 & -26\degr  32\arcmin  4\farcs03 & 4921 & 31.4 & 17.4 & ? &  & 0.21 & filled \Halpha{} absorption & 98/N11 (EC) & 2 \\
NGC 6218 & 16\nfh  47\nfm  10\fs59 & -1\degr  56\arcmin  40\farcs04 & 24182 & 55.8 & 20.3 & yes & ? &  & \Halpha{} emission &  &  \\
NGC 6218 & 16\nfh  47\nfm  12\fs31 & -1\degr  57\arcmin  18\farcs13 & 13339 & 36.5 & 19.6 & yes & ? &  & filled \Halpha{} absorption &  &  \\
NGC 6218 & 16\nfh  47\nfm  12\fs42 & -1\degr  57\arcmin  8\farcs41 & 13226 & 29.7 & 19.4 & ? & yes &  & variable, filled \Halpha{} absorption &  &  \\
NGC 6218 & 16\nfh  47\nfm  13\fs57 & -1\degr  57\arcmin  26\farcs47 & 12224 & 33.1 & 19.4 & ? & yes &  & variable \Halpha{}, \Hbeta{} emission &  &  \\
NGC 6218 & 16\nfh  47\nfm  15\fs71 & -1\degr  56\arcmin  46\farcs78 & 19958 & 24.4 & 20.8 & ? &  & 0.04 & broad \Halpha{} and \Hbeta{} emission & CX1 (CV) & 11 \\
NGC 6218 & 16\nfh  47\nfm  16\fs50 & -1\degr  57\arcmin  23\farcs75 & 9466 & 45.3 & 20.9 & yes &  &  & variable \Halpha{}, \Hbeta{} emission &  &  \\
NGC 6218 & 16\nfh  47\nfm  16\fs65 & -1\degr  56\arcmin  56\farcs24 & 9431 & 37.1 & 18.7 & ? & yes &  & variable, filled \Halpha{} absorption &  &  \\
NGC 6266 & 17\nfh  1\nfm  11\fs68 & -30\degr  6\arcmin  41\farcs92 & 4201 & 16.2 & 18.9 & yes & yes &  & variable \Halpha{} emission &  &  \\
NGC 6266 & 17\nfh  1\nfm  12\fs05 & -30\degr  6\arcmin  44\farcs59 & 708 & 10.8 & 15.5 & yes & no &  & variable, filled \Halpha{} absorption & V182 (RRLyr) & 2 \\
NGC 6266 & 17\nfh  1\nfm  12\fs16 & -30\degr  7\arcmin  2\farcs63 & 392 & 15.6 & 16.3 & yes & ? &  & variable, filled \Halpha{} absorption & V156 (RRLyr) & 2 \\
NGC 6266 & 17\nfh  1\nfm  12\fs54 & -30\degr  6\arcmin  22\farcs03 & 3208 & 27.6 & 18.4 & yes &  &  & very strong \Halpha{}, \Hbeta{} emission, misclassified? & V13 (RRLyr) & 2 \\
NGC 6266 & 17\nfh  1\nfm  12\fs77 & -30\degr  6\arcmin  45\farcs05 & 5 & 4.3 & 12.8 & yes & yes &  & very strong \Halpha{}, \Hbeta{} emission, misclassified? & V190? (RRLyr) & 2 \\
NGC 6266 & 17\nfh  1\nfm  12\fs83 & -30\degr  6\arcmin  46\farcs78 & 2106 & 2.7 & 17.5 & yes & ? &  & variable, filled \Halpha{} absorption & V188 (RRLyr) & 2 \\
NGC 6266 & 17\nfh  1\nfm  12\fs86 & -30\degr  6\arcmin  54\farcs94 & 6 & 5.6 & 14.0 & yes & yes &  & variable, filled \Halpha{} absorption, bad model? & V146 (CpII) & 2 \\
NGC 6266 & 17\nfh  1\nfm  12\fs94 & -30\degr  7\arcmin  21\farcs44 & 842 & 32.1 & 16.5 & yes & ? &  & variable, filled \Halpha{} absorption & V139 (RRLyr) & 2 \\
NGC 6266 & 17\nfh  1\nfm  12\fs98 & -30\degr  6\arcmin  19\farcs30 & 63 & 30.2 & 14.9 & no &  &  & variable \Halpha{}, \Hbeta{} emission & V239 (L) & 2 \\
NGC 6266 & 17\nfh  1\nfm  13\fs18 & -30\degr  6\arcmin  50\farcs28 & 1772 & 5.1 & 17.8 & yes & yes &  & variable \Halpha{} emission & M62-VLA1 (BH) & 12 \\
NGC 6266 & 17\nfh  1\nfm  13\fs41 & -30\degr  5\arcmin  59\farcs07 & 12285 & 51.0 & 19.4 & no & yes &  & \Halpha{} emission &  &  \\
NGC 6266 & 17\nfh  1\nfm  13\fs55 & -30\degr  6\arcmin  57\farcs45 & 222 & 12.7 & 15.9 & yes &  &  & variable, filled \Halpha{} absorption & V163 (RRLyr) & 2 \\
NGC 6266 & 17\nfh  1\nfm  13\fs81 & -30\degr  7\arcmin  9\farcs05 & 19 & 23.6 & 13.3 & yes & ? &  & variable \Halpha{} emission &  &  \\
NGC 6293 & 17\nfh  10\nfm  9\fs85 & -26\degr  34\arcmin  57\farcs61 & 152 & 5.1 & 16.0 & yes &  &  & variable, filled \Halpha{} absorption &  &  \\
NGC 6388 & 17\nfh  36\nfm  14\fs48 & -44\degr  43\arcmin  20\farcs75 & 250190 & 55.4 & 15.8 & yes & yes &  & variable \Halpha{}, \Hbeta{} emission & V18 (W Vir) & 2 \\
NGC 6388 & 17\nfh  36\nfm  15\fs41 & -44\degr  44\arcmin  14\farcs90 & 133353 & 20.6 & 14.8 & yes & ? &  & variable \Halpha{}, \Hbeta{} emission & V95 (SR) & 2 \\
NGC 6388 & 17\nfh  36\nfm  16\fs08 & -44\degr  44\arcmin  9\farcs51 & 128054 & 12.3 & 14.6 & yes & no &  & variable \Halpha{}, \Hbeta{} emission & V88 (SR) & 2 \\
NGC 6388 & 17\nfh  36\nfm  16\fs62 & -44\degr  44\arcmin  0\farcs71 & 233113 & 9.6 & 17.0 & yes & yes &  & \Halpha{} emission in single spectrum &  &  \\
NGC 6388 & 17\nfh  36\nfm  17\fs09 & -44\degr  44\arcmin  0\farcs53 & 229862 & 7.4 & 15.3 & yes & ? &  & variable \Halpha{}, \Hbeta{} emission & V98 (L) & 2 \\
NGC 6388 & 17\nfh  36\nfm  17\fs20 & -44\degr  44\arcmin  14\farcs85 & 119418 & 7.1 & 13.6 & yes & yes &  & variable \Halpha{}, \Hbeta{} emission & V73 (W Vir) & 2 \\
NGC 6388 & 17\nfh  36\nfm  17\fs29 & -44\degr  44\arcmin  18\farcs76 & 119389 & 11.0 & 12.8 & yes &  &  & variable, filled \Halpha{} absorption & V80 (RV Tau) & 2 \\
NGC 6388 & 17\nfh  36\nfm  17\fs47 & -44\degr  44\arcmin  17\farcs50 & 118083 & 10.1 & 16.6 & yes &  &  & \Halpha{}, \Hbeta{} emission & V69 (W Vir) & 2 \\
NGC 6388 & 17\nfh  36\nfm  17\fs60 & -44\degr  44\arcmin  16\farcs12 & 117361 & 9.2 & 15.3 & yes & ? &  & variable \Halpha{}, \Hbeta{} emission & V72 (W Vir) & 2 \\
NGC 6388 & 17\nfh  36\nfm  18\fs42 & -44\degr  44\arcmin  19\farcs39 & 111469 & 17.2 & 13.6 & yes & ? &  & variable \Halpha{}, \Hbeta{} emission & V2 (Mira Ceti) & 2 \\
NGC 6441 & 17\nfh  50\nfm  11\fs14 & -37\degr  3\arcmin  53\farcs05 & 1000228 & 53.1 & 15.0 & ? &  &  & variable \Halpha{}, \Hbeta{} emission & V9 (L) & 2 \\
NGC 6441 & 17\nfh  50\nfm  12\fs18 & -37\degr  3\arcmin  12\farcs45 & 135507 & 12.7 & 14.2 & yes & yes &  & variable \Halpha{}, \Hbeta{} emission & V127 (W Vir) & 2 \\
NGC 6441 & 17\nfh  50\nfm  12\fs42 & -37\degr  2\arcmin  52\farcs62 & 260896 & 14.7 & 16.0 & ? &  &  & \Halpha{}, \Hbeta{} emission, contaminated? &  &  \\
NGC 6441 & 17\nfh  50\nfm  12\fs64 & -37\degr  3\arcmin  12\farcs13 & 131318 & 8.5 & 15.1 & ? &  &  & variable \Halpha{} emission & V126 (W Vir) & 2 \\
NGC 6441 & 17\nfh  50\nfm  12\fs72 & -37\degr  2\arcmin  38\farcs05 & 258272 & 27.4 & 19.0 & ? &  &  & spectral changes, \Halpha{}, \Hbeta{} and Ca II triplett emission & V135 (L) & 2 \\
NGC 6441 & 17\nfh  50\nfm  13\fs69 & -37\degr  3\arcmin  16\farcs36 & 121826 & 13.5 & 18.7 & ? &  &  & \Halpha{}, \Hbeta{} emission, contaminated? & V139? (SR) & 2 \\
NGC 6441 & 17\nfh  50\nfm  15\fs63 & -37\degr  3\arcmin  7\farcs60 & 104171 & 30.9 & 15.3 & ? & ? &  & variable \Halpha{}, \Hbeta{} emission &  &  \\
NGC 6441 & 17\nfh  50\nfm  15\fs76 & -37\degr  2\arcmin  16\farcs45 & 231739 & 58.5 & 14.0 & ? & ? &  & variable \Halpha{}, \Hbeta{} emission & V6 (W Vir) & 2 \\
NGC 6441 & 17\nfh  50\nfm  16\fs29 & -37\degr  2\arcmin  40\farcs65 & 227201 & 45.9 & 16.6 & ? &  &  & variable \Halpha{}, \Hbeta{} emission & V2 (Mira Ceti) & 2 \\
NGC 6441 & 17\nfh  50\nfm  17\fs23 & -37\degr  3\arcmin  50\farcs01 & 91313 & 67.1 & 16.4 & yes &  &  & variable \Halpha{}, \Hbeta{} emission & V1 (Mira Ceti) & 2 \\
NGC 6522 & 18\nfh  3\nfm  32\fs50 & -30\degr  2\arcmin  3\farcs02 & 80 & 19.7 & 13.2 & yes &  &  & variable, filled \Halpha{} absorption &  &  \\
NGC 6522 & 18\nfh  3\nfm  35\fs42 & -30\degr  2\arcmin  6\farcs47 & 88 & 18.7 & 14.3 & yes & ? &  & variable \Halpha{} emission &  &  \\
NGC 6541 & 18\nfh  8\nfm  0\fs99 & -43\degr  43\arcmin  9\farcs53 & 44127 & 21.7 & 18.5 & no & yes & 0.05 & variable, filled \Halpha{} absorption & (SSG) &  \\
NGC 6541 & 18\nfh  8\nfm  2\fs22 & -43\degr  42\arcmin  58\farcs15 & 40733 & 4.8 & 17.4 & yes & yes & 0.18 & broad \Halpha{} emission & (RS) &  \\
NGC 6541 & 18\nfh  8\nfm  3\fs60 & -43\degr  43\arcmin  4\farcs50 & 36939 & 17.3 & 17.9 & yes & no & 0.10 & variable, filled \Halpha{} absorption & (SSG) &  \\
NGC 6656 & 18\nfh  36\nfm  21\fs68 & -23\degr  53\arcmin  35\farcs07 & 68961 & 52.2 & 15.7 & yes & yes & 0.29 & variable, filled \Halpha{} absorption & Source 33? (RS Cvn?) & 13 \\
NGC 6656 & 18\nfh  36\nfm  21\fs78 & -23\degr  54\arcmin  13\farcs36 & 68665 & 29.8 & 13.4 & yes & yes &  & variable \Halpha{} absorption & V24 (W Vir) & 2 \\
NGC 6656 & 18\nfh  36\nfm  23\fs79 & -23\degr  54\arcmin  18\farcs38 & 37810 & 2.3 & 13.7 & yes & ? &  & variable, filled \Halpha{} absorption &  &  \\
NGC 6656 & 18\nfh  36\nfm  24\fs05 & -23\degr  54\arcmin  29\farcs60 & 37211 & 12.6 & 10.7 & yes & no &  & variable filled \Halpha{} absorption & V35 (SR) & 2 \\
NGC 6656 & 18\nfh  36\nfm  24\fs10 & -23\degr  54\arcmin  33\farcs27 & 36997 & 16.3 & 18.2 & yes & ? &  & filled \Halpha{} absorption &  &  \\
NGC 6656 & 18\nfh  36\nfm  24\fs20 & -23\degr  54\arcmin  10\farcs32 & 63613 & 7.7 & 20.0 & ? & yes & 0.49 & noisy, broad \Halpha{} emission, multiple spectral components? &  &  \\
NGC 6656 & 18\nfh  36\nfm  24\fs69 & -23\degr  54\arcmin  35\farcs69 & 35776 & 21.3 & 18.9 & yes &  &  & broad \Halpha{}, \Hbeta{}, He emission & CV1 (CV) & 2 \\
NGC 6681 & 18\nfh  43\nfm  10\fs38 & -32\degr  17\arcmin  3\farcs09 & 39915 & 41.4 & 20.6 & no & ? &  & variable \Halpha{}, \Hbeta{} emission &  &  \\
NGC 6681 & 18\nfh  43\nfm  10\fs78 & -32\degr  17\arcmin  54\farcs57 & 21588 & 34.0 & 13.7 & yes & no &  & variable, filled \Halpha{} absorption &  &  \\
NGC 6681 & 18\nfh  43\nfm  11\fs78 & -32\degr  17\arcmin  55\farcs22 & 19706 & 26.6 & 22.7 & yes &  & 0.17 & broad \Halpha{}, \Hbeta{} and He emission & (CV) &  \\
NGC 6752 & 19\nfh  10\nfm  55\fs55 & -59\degr  59\arcmin  17\farcs40 & 14153 & 29.0 & 17.7 & yes &  & 0.49 & filled \Halpha{} absorption & CX19 (AB?) & 14, 15 \\
NGC 6752 & 19\nfh  10\nfm  56\fs00 & -59\degr  59\arcmin  37\farcs37 & 13797 & 44.0 & 19.6 & yes &  & 0.12 & broad \Halpha{}, \Hbeta{} and He emission & CX2 (CV) & 14 \\
NGC 7078 & 21\nfh  29\nfm  56\fs58 & 12\degr  10\arcmin  40\farcs19 & 186117 & 46.7 & 14.6 & yes & ? &  & filled \Halpha{} absorption &  &  \\
NGC 7078 & 21\nfh  29\nfm  57\fs83 & 12\degr  10\arcmin  0\farcs50 & 176122 & 7.3 & 16.0 & yes & yes &  & variable \Halpha{} absorption & V131 (RRLyr) & 2 \\
NGC 7078 & 21\nfh  29\nfm  57\fs84 & 12\degr  9\arcmin  51\farcs99 & 95810 & 11.6 & 15.9 & yes & yes &  & variable, filled \Halpha{} absorption & V130 (RRLyr) & 2 \\
NGC 7078 & 21\nfh  29\nfm  58\fs31 & 12\degr  10\arcmin  2\farcs80 & 172701 & 1.6 & 15.5 & yes &  & 0.13 & P Cygni of H, He & V125 (LMXB) & 2 \\
NGC 7078 & 21\nfh  29\nfm  59\fs15 & 12\degr  10\arcmin  7\farcs20 & 165986 & 13.5 & 13.6 & yes &  &  & variable \Halpha{}, \Hbeta{} emission & V86 (W Vir) & 2 \\
NGC 7078 & 21\nfh  29\nfm  59\fs43 & 12\degr  10\arcmin  21\farcs89 & 164189 & 26.3 & 19.1 & yes &  &  & variable \Halpha{} emission &  &  \\
NGC 7078 & 21\nfh  29\nfm  59\fs47 & 12\degr  9\arcmin  43\farcs80 & 83983 & 24.2 & 14.8 & yes & yes &  & variable \Halpha{} emission &  &  \\
NGC 7078 & 21\nfh  30\nfm  0\fs62 & 12\degr  10\arcmin  1\farcs35 & 154446 & 33.6 & 13.0 & yes & ? &  & emission from \Halpha{} wings &  &  \\
NGC 7089 & 21\nfh  33\nfm  23\fs64 & 0\degr  48\arcmin  59\farcs89 & 179324 & 56.0 & 14.5 & yes & ? &  & variable, filled \Halpha{} absorption &  &  \\
NGC 7089 & 21\nfh  33\nfm  23\fs84 & 0\degr  49\arcmin  12\farcs91 & 178318 & 48.8 & 12.7 & yes & ? &  & variable \Halpha{}, \Hbeta{} emission & V5 (W Vir) & 2 \\
NGC 7089 & 21\nfh  33\nfm  23\fs98 & 0\degr  49\arcmin  12\farcs48 & 177434 & 46.9 & 19.8 & yes &  &  & variable \Halpha{} emission &  &  \\
NGC 7089 & 21\nfh  33\nfm  24\fs34 & 0\degr  49\arcmin  9\farcs46 & 175600 & 42.6 & 16.3 & yes & ? &  & variable, filled \Halpha{} absorption &  &  \\
NGC 7089 & 21\nfh  33\nfm  24\fs93 & 0\degr  48\arcmin  43\farcs37 & 171867 & 51.1 & 16.2 & yes & yes &  & variable, filled \Halpha{} absorption & V51 (RRLyr) & 2 \\
NGC 7089 & 21\nfh  33\nfm  25\fs09 & 0\degr  48\arcmin  49\farcs44 & 170746 & 44.8 & 13.1 & yes & no &  & weak emission from \Halpha{} wings &  &  \\
NGC 7089 & 21\nfh  33\nfm  25\fs86 & 0\degr  49\arcmin  4\farcs93 & 165509 & 25.6 & 18.4 & yes & yes &  & variable, filled \Halpha{} absorption &  &  \\
NGC 7089 & 21\nfh  33\nfm  26\fs79 & 0\degr  48\arcmin  33\farcs64 & 219411 & 50.2 & 16.0 & yes & yes &  & variable, filled \Halpha{} absorption & V22 (RRLyr) & 2 \\
NGC 7089 & 21\nfh  33\nfm  27\fs52 & 0\degr  49\arcmin  59\farcs64 & 81620 & 36.7 & 13.3 & yes & yes &  & variable \Halpha{}, \Hbeta{} emission & V6 (W Vir) & 2 \\
NGC 7089 & 21\nfh  33\nfm  27\fs88 & 0\degr  49\arcmin  53\farcs45 & 79550 & 32.5 & 12.7 & yes & ? &  & emission from \Halpha{} wings &  &  \\
NGC 7099 & 21\nfh  40\nfm  20\fs96 & -23\degr  10\arcmin  43\farcs27 & 51191 & 16.5 & 18.0 & yes & yes &  & variable \Halpha{} absorption & (SSG) &  \\
NGC 7099 & 21\nfh  40\nfm  21\fs80 & -23\degr  10\arcmin  50\farcs55 & 26939 & 5.4 & 14.8 & yes & yes &  & variable, filled \Halpha{} absorption & V15 (RRLyr) & 2 \\
NGC 7099 & 21\nfh  40\nfm  21\fs80 & -23\degr  10\arcmin  9\farcs11 & 49462 & 38.6 & 11.7 & yes & ? &  & emission from \Halpha{} wings &  &  \\
NGC 7099 & 21\nfh  40\nfm  22\fs22 & -23\degr  10\arcmin  47\farcs46 & 25532 & 1.4 & 16.1 & yes & yes & 0.28 & variable \Halpha{} emission, between RGB and HB & A2 (?) & 16 \\
NGC 7099 & 21\nfh  40\nfm  22\fs95 & -23\degr  10\arcmin  49\farcs60 & 23423 & 11.7 & 20.3 & yes &  & 0.12 & broad \Halpha{} and \Hbeta{} emission & (CV) &  \\
\hline
\label{tbl:results}
\end{longtable}
\tablefoot{
\tablefoottext{a}{F625W.}
AB: chromospherically active binary, 
BL Her: BL Herculis pulsating variable,
CpII: Type II Cepheid pulsating variable,
CV: cataclysmic variable,
EA: eclipsing semi-detached binary (Algol-type), 
EC: eclisping contact binary,
EW: eclipsing low mass contact binary,
L: slow irregular variable, 
LMXB: low-mass X-ray binary,
LPV: long-period variable,
MSP: millisecond pulsar, 
SR: semiregular variable,
SSG: sub-subgiant,
RS: red straggler,
RS Cvn: close eruptive binary (RS Canum Venaticorum),
W Vir: W Virginis pulsating variable

\tablebib{
(1)~\citet{heinke_deep_2005}; (2)~\citet{clement_variable_2001}; (3)~\citet{albrow_frequency_2001}; (4)~\citet{kunder_variable_2013}; (5)~\citet{kaluzny_clusters_2016}; (6)~\citet{lebzelter_long-period_2016}; (7)~\citet{cool_hst/acs_2013}; (8)~\citet{heinke_chandra_2003}; (9)~\citet{dieball_far-ultraviolet_2010}; (10)~\citet{bassa_x-ray_2004}; (11)~\citet{lu_x-ray_2009-1}; (12)~\citet{chomiuk_radio-selected_2013}; (13)~\citet{webb_x-ray_2004}; (14)~\citet{pooley_optical_2002-1}; (15)~\citet{lugger_identification_2017}; (16)~\citet{lugger_chandra_2007}.}
}
\end{landscape}
\end{appendix}

%
%
\end{document}